\PassOptionsToPackage{a4paper, margin=1in}{geometry}

\documentclass[pdflatex,sn-mathphys-num]{sn-jnl}

\usepackage{graphicx}%
\usepackage{multirow}%
\usepackage[utf8]{inputenc}
\usepackage{amsmath,amssymb,amsfonts}%
\usepackage{amsthm}%
\usepackage{mathrsfs}%
\usepackage[title]{appendix}%
\usepackage{xcolor}%
\usepackage{textcomp}%
\usepackage{manyfoot}%
\usepackage{booktabs}%
\usepackage{algorithm}%
\usepackage{algorithmicx}%
\usepackage{algpseudocode}%
\usepackage{listings}%
\usepackage{makecell}
\usepackage{caption}

\usepackage[numbers]{natbib}
\usepackage[version=4]{mhchem} 
\usepackage{float}
\usepackage{xr}
\usepackage{tikz}
\usepackage{geometry}

\usetikzlibrary{arrows,positioning,shapes,decorations.pathmorphing,arrows.meta,math}
\usetikzlibrary{decorations.pathreplacing,patterns,intersections,backgrounds,calc}
\usepgflibrary{shapes.geometric}

\def\de{\mathrm{d}}

\begin{document}

\title[Article Title]{The Value of Patience in Online Grocery Shopping}


\author[1,2,6]{\fnm{Javad} \sur{Eshtiyagh}} \email{javade@mit.edu}
\author*[1,6]{\fnm{Pei} \sur{Zhao}}\email{peizhao@mit.edu}
\author[3,6]{\fnm{Federico}\sur{Librino}}\email{federico.librino@iit.cnr.it}
\author[3]{\fnm{Giovanni} \sur{Resta}}\email{g.resta@iit.cnr.it}
\author[1,3]{\fnm{Paolo} \sur{Santi}}\email{psanti@mit.edu}
\author[1]{\fnm{Martina} \sur{Mazzarello}}\email{mmazz@mit.edu}
\author[4]{\fnm{Akanksha} \sur{Khurd}}\email{askhurd@iu.edu}
\author[4]{\fnm{Santo} \sur{Fortunato}}\email{santo@iu.edu}
\author[1,5]{\fnm{Carlo} \sur{Ratti}}\email{ratti@mit.edu}

\affil[1]{Senseable City Lab, Massachusetts Institute of Technology, Cambridge, MA, USA}
\affil[2]{Transport Strategy Centre, Imperial College London, London, UK}
\affil[3]{Instituto di Informatica e Telematica del CNR, Pisa, Italy}
\affil[4]{Luddy School of Informatics, Computing, and Engineering, Indiana University, Bloomington, USA}
\affil[5]{Department ABC, Politecnico di Milano, Milano, Italy}
\affil[6]{These authors contributed equally to this work}

\abstract{
Since the COVID-19 pandemic, online grocery shopping has rapidly reshaped consumer behavior worldwide, fueled by ever-faster delivery promises aimed at maximizing convenience. Yet, this growth has also substantially increased urban traffic congestion, emissions, and pollution \citep{erhardt2019transportation,kang2021low,kikstra2021climate,hoehne2017greenhouse,guo2017correlations}. Despite extensive research on urban delivery optimization, little is known about the trade-off between individual convenience and these societal costs. In this study, we investigate the value of marginal extensions in delivery times—termed customer patience—in mitigating the traffic burden caused by grocery deliveries. We first conceptualize the problem and presents a mathematical model that highlight a convex relationship between patience and traffic congestion. The theoretical predictions are confirmed by an extensive, network-science based analysis leveraging two large-scale datasets encompassing over 8 million grocery orders in Dubai. Our findings reveal that allowing just five additional minutes in delivery time reduces daily delivery mileage by approximately 30\% and life-cycle \ce{CO2} emissions by 20\%. Beyond ten minutes of added patience, however, marginal benefits diminish significantly. These results highlight that modest increases in consumer patience can deliver substantial gains in traffic reduction and sustainability, offering a scalable strategy to balance individual convenience with societal welfare in urban delivery systems.

}

\maketitle

\section{Main}

The rapid growth of online shopping has driven a significant expansion of urban delivery services, a trend accelerated by the COVID-19 pandemic \citep{dablanc2017rise}. In 2024, e-commerce accounted for approximately 20\% of retail sales globally, with nearly one-third of the world’s population engaging in it \citep{salesshare, usershare}.

While urban delivery has greatly enhanced consumer convenience—especially through increasingly faster delivery times—the continued expansion of delivery fleets has introduced significant challenges. These include increased traffic congestion \citep{erhardt2019transportation}, rising carbon emissions \citep{kang2021low, kikstra2021climate}, and deteriorating urban air quality \citep{hoehne2017greenhouse, guo2017correlations}.

Several studies have explored strategies to mitigate the societal costs of urban delivery. On the supply side, logistics companies such as Amazon and UPS have focused on fleet optimization and shipment consolidation \citep{AmazonShipping,UPSShipping}. Parallel efforts have investigated the adoption of low-carbon technologies, including electric vehicles \citep{woody2022optimizing}, drones \citep{moadab2022drone}, and automated delivery systems \citep{boysen2021last}. More disruptive approaches, such as fleet mixing \citep{arslan2019crowdsourced,diao2024multi}, public transportation integration \citep{gao2024new}, crowd sourcing \citep{ghaderi2022integrated,dahle2019pickup,arslan2019crowdsourced}, and order-splitting \citep{arslan2019splitting}, have also been proposed. However, these strategies often require major infrastructure overhauls with resulting high upfront costs, which pose substantial barriers to widespread adoption \citep{li2025beyond, stolaroff2018energy}.

Interventions on the demand side—targeting consumer behavior—offer promising and underexplored opportunities for reducing emissions. In particular, one factor that has been largely overlooked is customer patience, or willingness to accept longer delivery times. Longer delivery windows allow for greater bundling of deliveries, thereby reducing total travel requirements. Here, we aim to characterize the universal relationship between delivery time flexibility and resulting traffic flows.

As a case study, we focus on real-time grocery delivery in Dubai, analyzing two large-scale grocery delivery datasets comprising over 8 million orders. We first approach the problem theoretically, presenting a conceptual model that relates system parameters to the probability of bundling two orders, and highlights a convex relationship between customer patience and traffic congestion. We then develop a network science-based framework to optimize both order bundling and vehicle allocation under real-time constraints that is capable of addressing the computational complexity issue that has previously hindered large-scale practical applications of such optimization. The results of the empirical analysis confirm the predictions of the theoretical model and reveal a nuanced relationship between customer patience, traffic congestion, and environmental costs.

\subsection*{Bundling and Dispatching Strategies} 

\begin{figure}[H]
    \hspace*{-1.2cm}
    \centering
    \includegraphics[width=1.1\linewidth]{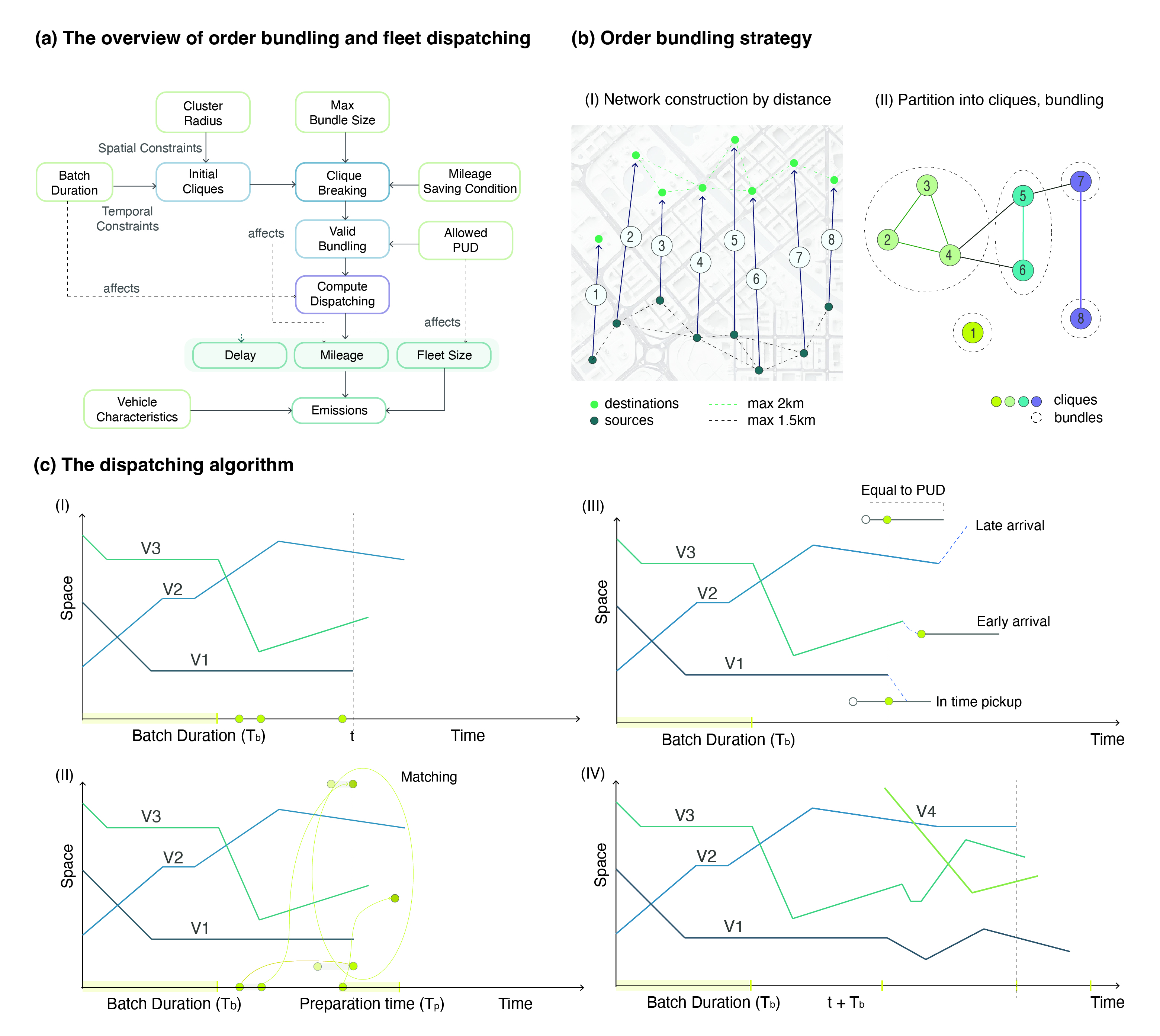}
    \caption{Overview of the developed order bundling and fleet dispatching framework. (a) Overall workflow. Orders are bundled based on spatial and temporal constraints (in light green except for vehicle characteristics), then dispatched to available vehicles. Key outputs, including delivery delay, mileage, fleet size, and emissions, were estimated. (b) Bundling strategy to form an order shareability network. (b-I) Orders that meet the predefined spatial and temporal proximity constraints in a batch are considered as bundling candidates; (b-II) These candidates form a shareability network, where nodes represent orders and cliques indicate feasible bundling opportunities. (c) The fleet dispatching algorithm. (c-I) At the end of a batch, orders issued during the batch are collected (and bundled) and vehicles idle positions are estimated; (c-II) Order ready times and the weighted order-vehicle matching are computed; (c-III) depending on the assignment and  PUD, vehicles able to arrive at the pickup point on time or earlier are assigned, and their next idle positions are updated; (c-IV) For orders unreachable in time by any vehicle, a new vehicle is generated at the pickup point (vehicle V4), thus increasing the fleet size.}
  
    \label{fig:overview}
\end{figure}

To estimate the societal benefits of customer patience in last-mile delivery, we propose an efficient strategy that integrates real-time order bundling with intelligent fleet dispatching. Figure \ref{fig:overview}(a) provides an overview of our methodological framework. This strategy relies on several key parameters: the batch duration ($T_{b}$), which defines the time window for collecting and potentially bundling orders; the maximum bundle size ($k$); the maximum pickup delay (PUD), defined as the maximum time an order can wait at the store before pickup; and the maximum delivery delay, measured relative to the time the order would have been delivered without bundling. Spatial proximity constraints are defined by distances between origin grocery stores ($d_v$) and between destination customers ($d_c$).

Under different parameter settings, the optimized strategy outputs delivery fleet size, total mileage, life-cycle emissions, and average delivery delay. Total mileage represents the total distance traveled by the delivery fleet to complete the bundled orders, including empty miles during repositioning between drop-offs and subsequent pickups. This total mileage serves as a proxy for the local impact of delivery services, particularly on traffic congestion and urban air pollution. Life-cycle emissions, on the other hand, account for the overall environmental costs associated with producing, operating, and disposing of the delivery vehicle fleet, and can be considered a proxy for the global environmental impact of the delivery service.

For order bundling, we extend the concept of a shareability network, previously proposed for ride-sharing \cite{Santi2014,Agatz2012}, by introducing a network-based approach to model bundling opportunities between grocery orders (see Figure \ref{fig:overview}(b)). More specifically, in the {\em order shareability network}, each order is represented as a node, and edges connect orders that meet predefined spatial and temporal proximity constraints within a batch duration. Two orders are considered potential candidates for bundling if they satisfy the requirements of vendor proximity constraints, customer proximity constraints, and temporal conditions, as illustrated by packages 2, 3, and 4 in Figure \ref{fig:overview}(b).

By partitioning the order shareability network into sub-cliques of maximum bundle size $k$, it is possible to minimize the number of bundled order deliveries while ensuring that $i)$ all orders are served and $ii)$ origin and destination proximity constraints are met. While finding the minimum clique cover (clique partitioning) in a network is computationally intractable, efficient heuristics with strong practical performance exist. A detailed explanation of the order bundling strategies is provided in the Methods section.

Building on the bundled orders (cliques) from the order shareability network partitioning, the dispatching strategy dynamically allocates delivery tasks to vehicles in the fleet, as illustrated in Figure \ref{fig:overview}(c). Using a batch-iterated process, bundled orders within a batch window are considered for vehicle assignment (c-I). Then, in (c-II), the ready times of all orders are determined. A bipartite network is constructed, linking vehicles to bundled orders to ensure timely pickups based on their last known coordinates (LKC). A minimum-weight matching algorithm is then applied to minimize the total mileage, with unassigned deliveries triggering additional vehicles as needed. This iterative approach ensures that deliveries are completed within the maximum allowed pickup delay (PUD) while optimizing fleet size and total mileage. This batch-based dispatching strategy approximates the minimum fleet size required to serve all orders, which would otherwise require full knowledge of all future orders \citep{Vazifeh2018}. This approximation substantially reduces the computational burden, making a real-time implementation of the proposed framework feasible. See Supplementary Note \ref{SI-note:ComputationTime} for further details on the algorithm’s computational time.

\subsection*{Conceptualizing the Relationship Between Patience and Bundling}

\begin{figure}[H]
 \centering
 \includegraphics[width=1\linewidth]{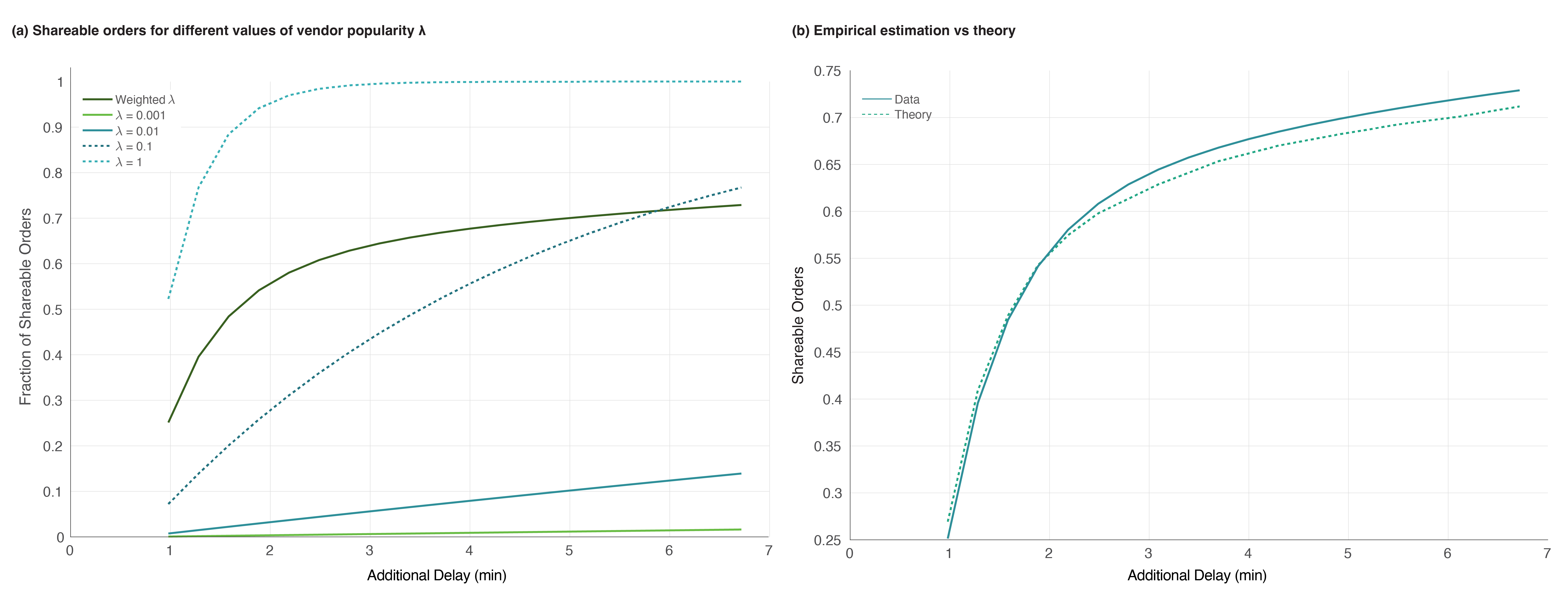}
 \caption{(a) Fraction of shareable orders for different values of vendor popularity $\lambda$ as a function of additional delay (patience $\theta$). (b) Estimated fraction shareable orders as a function of the customer patience based on theory and data.}
 \label{fig:Figure2_shapat_math}
\end{figure}

To provide a theoretical foundation for the trade-off between consumer patience and delivery efficiency, we begin with a simplified setting where at most two orders can be bundled together ($k=2$), and both orders must originate from the same vendor. As demonstrated in Supplementary Figure \ref{SI-fig:bundle_size_contours}, for short batch durations, which we focus on in this study, the case of bundles of size 2 is the most common bundling outcome based on the empirical grocery data from Dubai. Under this setting, the bundling opportunity is determined only by the temporal proximity of the orders and the spatial closeness between the two customers. Additionally, we assume a linear relationship between batch duration and average customer patience $\theta$ -- an assumption that is confirmed by empirical data (see details in Supplementary Note \ref{SI-note:TheoreticalModel}). Under these assumptions, we consider an order \emph{shareable} when it can potentially be bundled with at least one other order from the same vendor. This probability can be mathematically expressed as:

 \begin{equation}\label{eq:shareable_orders}
      P(\lambda, \theta) = 1 - \frac{2}{\lambda(w\theta+z)}\left(e^{-\frac{\lambda(w\theta+z)}{2}}-e^{-\lambda(w\theta+z)}\right).
 \end{equation}
 where $\theta$ is the average customer patience; $w$ and $z$ are parameters that characterize the linear relationship between the batch duration and the average customer patience; and $\lambda$ is the vendor popularity. As demonstrated in Figure \ref{fig:Figure2_shapat_math} (a), vendor popularity, defined as the average number of orders departing from that vendor per second, is a key determinant of the fraction of shareable orders. 
  

Equation \ref{eq:shareable_orders} holds for a vendor with a known popularity $\lambda$. To find the average shareability probability in a given city, the curve must be averaged across what we call the \emph{posterior} vendor popularity distribution, that is, the distribution of the popularity of the vendor of a randomly chosen order within the city. It can be shown that this distribution is expressed as $\lambda f(\lambda)/\hat\lambda$. Here, $f(\lambda)$ is the \emph{prior} popularity distribution, that is, the vendor popularity distribution of the considered city, while $\hat\lambda$ is the average vendor popularity. The general expression hence reads

 \begin{eqnarray}
  P(\theta) & = & 1 - \frac{2}{\hat\lambda\left(w\theta+z\right)}\mathbb{E}_\lambda\left[e^{-\frac{\lambda(w\theta+z)}{2}}-e^{-\lambda(w\theta+z)}\right] \nonumber\\
  & = & 1 - \frac{2}{\hat\lambda\left(w\theta+z\right)}\int_0^\infty\left(e^{-\frac{\lambda(w\theta+z)}{2}}-e^{-\lambda(w\theta+z)}\right)f(\lambda)\de\lambda.
 \end{eqnarray}

When the vendor popularity distribution and other relevant parameters are derived from Dubai data set, we obtain a theoretical curve of the relationship between customer patience $\theta$ and order shareability probability $P(\theta)$  that closely resembles the empirical curve obtained from data -- see Figure~\ref{fig:Figure2_shapat_math}(b). The theoretical curve closely approximates the data across the entire considered 1 to 7-minute span of the patience parameter $\theta$. The derived formula reveals a convex relationship between consumer patience and bundling opportunity: when average customer patience increases from 1 to 5 minutes, the fraction of shareable orders grows rapidly from less than 30\% to around 70\%; conversely, the increase in bundling opportunity is much slower when patience extends beyond five minutes. This convexity indicates that small increases in patience can lead to disproportionately large delivery optimization potential (shareability probability). We further translate this potential into quantitative societal impacts using the proposed bundling and dispatching strategies in the following section. 

It is important to note that the shareability probability $P(\theta)$ represents the potential for an order to be bundled, but actual bundling depends on the mutual compatibility of sharing opportunities across orders. Nevertheless, prior research on taxi ride sharing, including Vazifeh et al. \cite{Vazifeh2018}, has demonstrated that shareability probability is strongly correlated with realized sharing ratios. In the Methods section below and Supplementary Note \ref{SI-note:TheoreticalModel}, we provide further details demonstrating that this close correspondence also holds in the context of grocery delivery.


\subsection*{Empirical Results Based on Grocery Delivery Datasets}
We apply the proposed bundling and dispatching strategy to two large-scale e-commerce datasets in Dubai to empirically analyze the trade-off between customer patience and social cost. The datasets contain over 8 million online grocery delivery records obtained from two leading grocery delivery companies (referred to as \textit{Data Provider 1} and \textit{Data Provider 2}) in the UAE. Delivery data from Data Provider 1 and 2 were collected in 2023 and in 2022, respectively. Each record includes detailed information such as customer locations, order placement time, and delivery time. Detailed descriptions of the datasets can be found in the Methods section.

Utilizing the proposed bundling strategy, we first compared the empirical simulation results with the theoretical derivations of bundling at most two deliveries ($k=2$). As shown in Supplementary Note \ref{SI-note:TheoreticalModel}, we found the theoretical and empirical results have highly consistent estimations of the influence of patience on saving mileage with a calculated $R^2$ larger than 0.99. Both results confirm the convex nature of the tradeoff between consumer patience and mileage savings at $k=2$: the fraction of saved mileage rises rapidly from 6\% to around 21\% when patience increases from 1 minute to 5 minutes, and then flattens beyond additional patience over 5 minutes.

Expanding from the simplest case with a maximum bundling size of $k=2$, we further quantified the trade-off between customer patience and societal impact with larger bundling sizes using the empirical dataset, which better reflects real-world conditions. For societal impacts, we evaluated total fleet mileage as an indicator of local impacts, and life-cycle \ce{CO2} emissions as an indicator of global impacts.

\begin{figure}[H]
    \centering
    \includegraphics[width=1\linewidth]{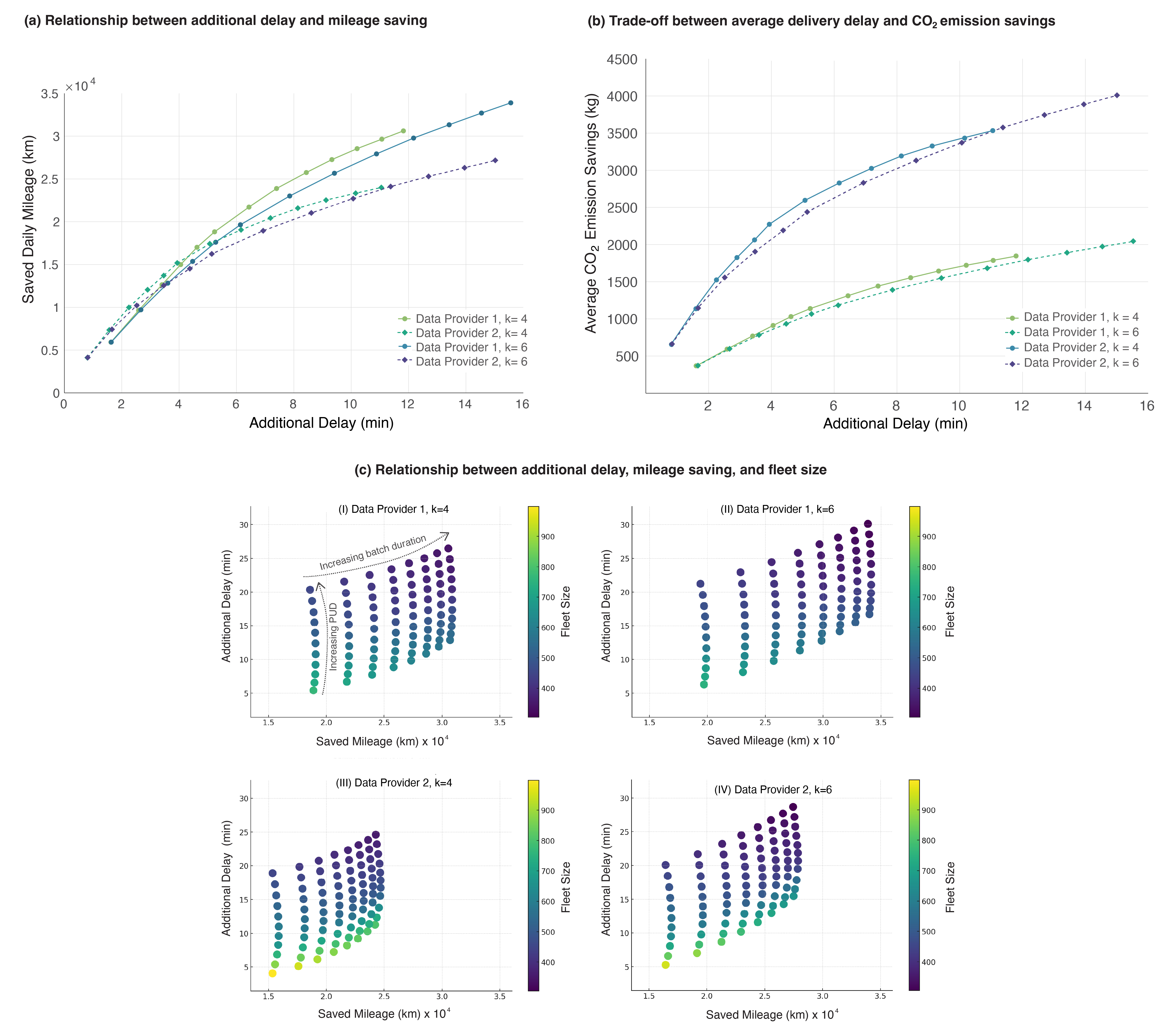}
    \caption{The tradeoff between average delivery delay (patience $\theta$) with (a) total mileage savings, (b) life-cycle \ce{CO2} emissions, and (c) fleet size.}
    \label{fig:Figure3_tradeoff}
\end{figure}

Figure \ref{fig:Figure3_tradeoff}(a) and (c) show the relationship between additional delay (patience $\theta$) and daily total mileage savings and fleet size changes under various batch durations values. Consistent with the simplified case ($k=2$), the relationship between average delay and mileage savings for $k=4$ or $k=6$ remains concave, indicating substantial benefits at small increases in patience, followed by diminishing marginal returns in mileage savings with longer delays. For instance, with an additional delay of 5 minutes, mileage savings reach approximately 13,000 km at $k=6$ for provider 2. However, increasing the delay from 10 to 15 minutes yields only about 5,000 km of additional mileage savings.  

From a life-cycle perspective for the delivery fleet, mileage savings and fleet size change from the strategy would lead to \ce{CO2} emission reductions for the delivery fleets. Therefore, based on the simulation results and GREET model~\cite{GREET} with localized emission factors, we assessed the life-cycle \ce{CO2} emission reductions from the proposed order bundling and dispatching strategy. As shown in Figure \ref{fig:Figure3_tradeoff}(b), there exists a similar trade-off between the emission reductions and average additional delay (patience $\theta$). Benefiting from both lower delivery mileage and decreased fleet size, as illustrated in \ref{fig:Figure3_tradeoff}(c), with an additional customer patience of 5 minutes in the case of $k=4$, it is expected to have 20\% life-cycle \ce{CO2} emission reduction for the delivery fleets. The tradeoff between patience and emission reduction presents similar convexity compared with that between patience and mileage savings. The fraction of life-cycle \ce{CO2} emission reduction increased rapidly when customers have an extra 5 minutes of patience, while the reduction percentage flattens with patience increasing beyond 5 minutes.

Additionally, we examine the influence of the maximum bundle size $k$ on the simulated results. As shown in Supplementary Figures \ref{SI-fig:k_influence_delay_and_mileage} and \ref{SI-fig:bundle_size_distribution}, increasing the bundle size to have four maximum bundled orders ($k=4$) yields a significant reduction in total emissions compared to no bundling scenario ($k=1$). However, beyond $k=4$, the marginal environmental benefits of further increasing the bundle size diminish. When limiting additional delay to less than 15 minutes, high values of $k$, i.e. $k>5$, do not lead to higher mileage savings while substantially increasing additional delivery delays.

\begin{figure}[H]
    \centering
    \includegraphics[width=1\linewidth]{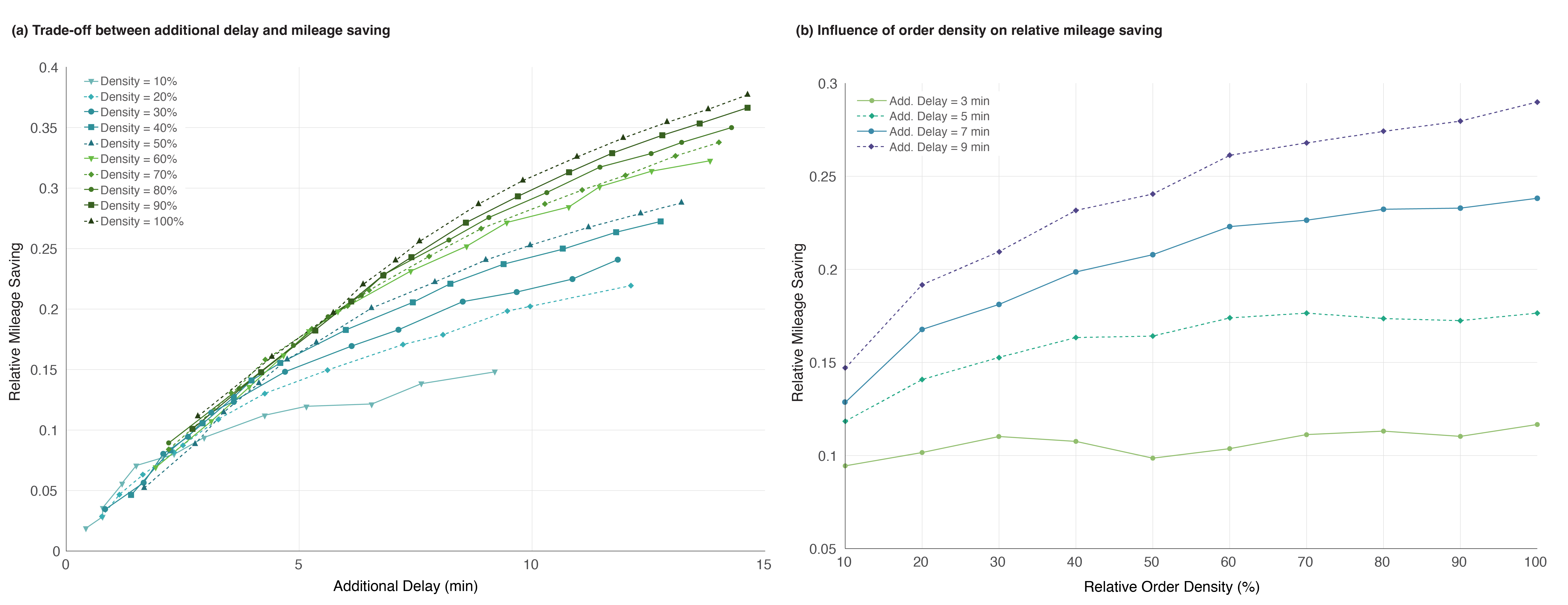}
    \caption{The influence of order density on the tradeoff between average delivery delay (customer patience $\theta$), and total mileage savings.}
    \label{fig:Figure4_density}
\end{figure}

To assess the robustness of the trade-off in different cities which might have different order density, we evaluate mileage savings and average delivery delay on the day with the largest number of orders by randomly sub-sampling the orders in 10\% percentile batches. As shown in Figure \ref{fig:Figure4_density}, the convexity of patience–mileage/emission trade-off remains consistent across all order densities: marginal mileage and emissions reduction benefits diminish as the patience increases. Higher order densities enable more effective bundling, which substantially increases the relative mileage/emission savings. For instance, in Figure \ref{fig:Figure4_density}(b), a 5-minute additional delay results in a 12\% reduction in \ce{CO2} emissions at 10\% order density, compared to an 18\% reduction at 100\% density. At the maximum daily order volume, the bundling strategy with $k=4$ achieves a 47\% mileage saving under a 10-minute delivery delay constraint. These results suggest that in scenarios with higher delivery demand—whether in the future or in different regions—the proposed strategy has greater potential to reduce social costs without requiring additional customer waiting time.

\subsection*{Discussion}

In the bundling and dispatching strategy, several parameters, such as batch duration, maximum allowed pickup delay (PUD), and bundling radius ($r$), influence outputs including total mileage savings, average delivery delay, and emissions. Therefore, a sensitivity analysis was conducted in Supplementary Note \ref{SI-note:SensitivityAnalysis} to quantify the influence of these parameters.

In this study, we find that a small compromise in consumer patience would lead to substantial environmental benefits of grocery delivery at both the local and global level. Both theoretically and empirically, we proved the convexity of the tradeoff between the patience and societal cost, where the marginal benefits decreased with increased patience. This highlights the importance of the small behavior change, i.e., a 5-minute delivery delay, in reducing the societal impact of urban delivery from optimized delivery practices.

In the broader context of the role of behavioral changes in combating climate change, our study contributes to a thorough characterization of the technical potential~\cite{Nielsen2024} of a simple consumer behavioral change—allowing slightly more flexibility in grocery delivery times—to mitigate the climate impact of express delivery. According to \cite{Nielsen2024}, the technical potential is the reduction in GHG emissions if all targeted individuals change their behaviour as intended, serving as a fundamental first step in identifying the most effective climate mitigation strategies~\cite{Nielsen2024}. In this context, our analysis can be interpreted as an assessment of the best-case scenario in emission reductions, should all consumers change their behavior as assumed in our study. 

Our study also evaluated the influence of behavior plasticity~\cite{Nielsen2024} in the context of urban grocery delivery, which is also a key factor used in prioritizing climate mitigation strategies~\cite{Nielsen2024}. The behavioral change considered in our analysis—a modest increase (e.g., 5 minutes) in consumer patience for receiving grocery items—would require low behavioral plasticity, as it involves only minimal changes to daily routines. Importantly, the convexity of the relationship between consumer patience and societal benefits indicates that eliminating the societal impact from urban grocery delivery does not require a huge change in customer behavior, as the marginal benefits diminished with the increase of patience. The high technical potential and low behavior plasticity demonstrate the high deployed opportunities of the proposed strategy and its effectiveness in reducing societal impacts \citep{dhal2025}.

Additionally, the optimization strategy proposed in this paper can provide direct cost advantages. By reducing both fleet size and total mileage traveled, the strategy improves operational efficiency without reducing revenue, as the number of delivered orders remains unchanged. We estimate that adopting the optimal bundling and dispatching strategy with a 10-minute increase in consumer patience could reduce delivery costs by approximately 13\% (see details in Supplementary Note \ref{SI-note:cost}), which delivery companies could use to incentivize customer patience. Therefore, the proposed strategy is likely to create a triple-win for delivery companies, customers, and cities.

Regarding generalizability, the proposed strategy and findings of this study can be extended to other cities and last-mile delivery scenarios. By focusing on estimating the technical potential of a consumer behavior change, we evaluate the potential of an intervention using the spatial and temporal characteristics of delivery orders, while accounting for fleet operational constraints. In analogous point-to-point transportation problems, it has been shown that sharing opportunities follow remarkably similar dynamics—primarily driven by demand density and traffic speed—across several cities \cite{Tachet2017}. This suggests that our findings are likely to generalize well to other urban contexts. From a global perspective, introducing a five-minute delay in online grocery deliveries could reduce carbon emissions equivalent to the annual absorption of 366 million trees in 2023 and 531 million trees in 2028. This reduction corresponds to a decrease in the annual social cost of carbon \cite{rennert2022comprehensive} by an estimated \$1.47 billion in 2023 and \$2.14 billion in 2028 (see Supplementary Note \ref{SI-note:globalprojection} for details). 

As for generalizing to other last-mile delivery services, while the framework developed in this study is tailored to the specific operational and logistical constraints of grocery delivery, it can be adapted to accommodate the constraints of other delivery services. This flexibility enables similar quantitative evaluations of the fundamental trade-off in a wide range of last-mile delivery settings.

\section*{Methods}
\subsection*{Data Overview}
We used two delivery datasets from Dubai. For privacy reasons, the e-commerce companies that provided the data for this study will remain unnamed and will be referred to as Provider 1 and Provider 2. The primary dataset used in this study for express deliveries is provided by Provider 1, one of the largest food delivery companies in the Middle East. The dataset spans from January 1, 2023, to December 31, 2023, capturing detailed information on customer orders over this period. It encompasses over 6 million grocery orders, 47 unique vendors, and over 2 million unique users. Each order record includes key attributes such as the locations of the vendor and consumer, order placement timestamps, delivery timestamps, and user IDs etc. The second dataset is obtained from Provider 2, which is one of the most popular grocery chains in the UAE. This dataset is comprised of similar variables as Provider 1 for all online orders in 2022. In this dataset, there are over 2 million orders submitted by approximately 1 million unique customers to 41 vendors.

While both Provider 1 and Provider 2 are treated similarly in this study, in reality, Provider 1 grocery orders represent express online deliveries, whereas Provider 2 orders are scheduled for next-day delivery. It should be noted that, due to the distribution of stores and customers, Provider 1 orders tend to cover longer distances, with a mean delivery distance of approximately 3.2 km, compared to 1.9 km for Provider 2 orders. Moreover, the Provider 1 dataset is denser, with each store fulfilling an average of 81,954 orders, compared to 51,915 orders per Provider 2 store. Additionally, Provider 1 users place an average of 2.7 orders per year, whereas Provider 2 users place an average of 1.97 orders. Further details and illustrations of the two datasets are provided in Supplementary Figures \ref{SI-fig:spatial_data_overview} and \ref{SI-fig:data_overview_graphs} and Supplementary Table \ref{SI-tab:data_overview_table}.

Given the size of the large dataset, most of the analysis is conducted on a representative sample consisting of four complete weeks from different seasons—January, April, August, and November. Additionally, as described, we apply our analysis to days with the minimum (7,329), first quartile (15,985), median (17,322), third quartile (18,987), and maximum (26,542) number of daily Provider 1 orders throughout the year. Lastly, when comparing Provider 1 and Provider 2 results, we downsampled the first dataset by approximately 43.9 percent to ensure comparability at similar order volumes.

\subsection*{Order bundling}
The bundling of express grocery packages is accomplished by applying a network-theoretical method based on clique partitioning to form bundles. In this approach, we construct an order shareability network where each order is represented as a node. Pairs of nodes corresponding to orders that could potentially be bundled are connected by an edge. 
Two orders are considered potential candidates for bundling if their vendors are within a specified distance $d_v$, their destinations (clients) are within a specified distance  $d_c$, and their ready-for-pickup times fall within the same batch, as shown in Figure \ref{fig:overview}. 

The resulting order shareability network is then partitioned into cliques with the goal of minimizing their number. Cliques are desirable because they implicitly enforce a locality criterion.
Indeed, requiring that the vendors (respectively, the clients) in a clique have a mutual distance of at most  $d_v$ (respectively,  $d_c$)
 implies that they can be circumscribed by a circle with a radius of 
$d_v/\sqrt{3}$ (respectively, $d_c/\sqrt{3}$), according to 
Jung's Theorem \citep{MR157289}.
Similarly, the time constraint in the construction of the shareability network implies that all the orders corresponding to a clique are ready for pickup within a time window not exceeding $T_B$.


By partitioning the order shareability network into cliques, we ensure that each single order is served as part of a bundle (clique) -- note that singleton cliques are allowed in the partitioning. Hence, by minimizing the number of cliques needed to partition the order shareability network -- a problem called {\em minimum clique cover}, we can determine the minimum number of bundles needed to serve all the orders.

In general, finding a minimum clique cover of a graph $G$ is NP-hard, as it is equivalent to graph coloring problem on the complement of $G$ \citep{Karp1972}.
However, polynomial-time heuristics do exist. Among these, we selected one that strikes a balance between producing good results and being easy to implement  \citep{Bhasker1991}, with a worst-case cost of $O(n^2)$, where $n$ is the number of orders to be bundled.


In the following, the function $\Phi_t(x,y)$ indicates the expected time needed to go from location $x$ to location $y$: it is obtained using the OSRM application, and adjusted to include the impact of vehicular traffic with hourly resolution. Similarly, $\Phi_d(x,y)$ indicates the road distance between the two locations $x$ and $y$.

Once we have partitioned the order shareability network into cliques, we then split the larger cliques to restrict the size of the bundles to the number of packages that a single delivery vehicle can transport ($k$), using a polynomial-time greedy heuristic. The division into bundles is designed so that each bundle is either a singleton or provides a reduction in mileage compared to the sum of individual deliveries of the orders within the bundle.

For example, for a bundle of two orders, with vendors locations $V_1$ and $V_2$ and respective client locations $C_1$ and $C_2$, we compare the sum of the lengths of the two original paths 
$d_o=\Phi_d(V_1,C_1)+\Phi_d(V_2,C_2)$ 
with the lengths of the bundled paths
\begin{align*} 
d_1&=\Phi_d(V_1,V_2)+\Phi_d(V_2,C_2)+\Phi_d(C_2,C_1)\,,\\ 
d_2&=\Phi_d(V_1,V_2)+\Phi_d(V_2,C_1)+\Phi_d(C_1,C_2)\,,\\
d_3&=\Phi_d(V_2,V_1)+\Phi_d(V_1,C_1)+\Phi_d(C_1,C_2)\,,\\
d_4&=\Phi_d(V_2,V_1)+\Phi_d(V_1,C_2)+\Phi_d(C_2,C_1)\,,
\end{align*}
and we accept the bundling only if $d_b<d_o$, where 
 $d_b=\min(d_1,d_2,d_3,d_4)$ is the length of the shortest bundled path.

In the presented results, we used `as the crow flies' distances to construct the order shareability network for efficiency. However, to accurately evaluate the mileage and emission reductions, we employed the Open Source Routing Machine (OSRM) to compute the source-destination paths, thus assessing the actual mileage advantage corresponding to a bundled delivery.

\subsection*{Dispatching}
The purpose of the dispatching operation is to assign each bundled order delivery to a vehicle of the fleet, provided that a given constraint on the delivery delay is matched.
In this study, we propose a framework that iteratively processes the deliveries in \emph{batches}. Besides being computationally viable, especially when thousands of orders are to be assigned to hundreds of vehicles, this approach better fits the investigated express delivery scenario.
The same approach can be adopted for single or bundled deliveries. We start by describing how the algorithm works in the former case, then we will detail how it is modified to account for bundled orders.

Each delivery $D_i$ is uniquely identified by the tuple $(\ell_v(i), \ell_c(i), t_o(i))$, where
\begin{itemize}
    \item $\ell_v(i)$ is the vendor location, where the item must be picked up;
    \item $\ell_c(i)$ is the customer location, where the item must be delivered;
    \item $t_o(i)$ is the order time, that is, the time instant when the order enters the system.
\end{itemize}
Since items usually require some time to be prepared, it is more practical to consider an alternative tuple $(\ell_v(i),\ell_c(i),t_r(i))$, where $t_r(i)$ is the time instant at which the item is ready to be picked up at the vendor location. The preparation time is, in general, different across orders; however, without loss of generality, in this work we consider it fixed by setting $t_r(i) = t_o(i) + T_p$, where $T_p$ is constant (5 minutes).
Notice that, even if the order is ready at time $t_r(i)$, the pickup may occur later, depending on vehicle availability. Hence, we define as $t_p(i)\geq t_r(i)$ the effective pickup time.
The difference $t_p(i)-t_r(i)$ is the pickup delay (PUD) of delivery $D_i$.
Finally, we call $t_d(i) = t_p(i) + \Phi_t(\ell_v(i), \ell_c(i))$ the delivery time, which is obtained by adding to the pickup time the travel time required to go from $\ell_v(i)$ to $\ell_c(i)$.

For each vehicle $V_k \in \mathcal{V}$, where $\mathcal{V}$ is the set of vehicles in the fleet, at any time $t$ we can define its \emph{last known coordinates} (LKC) as the pair $(\sigma_k, \tau_k)$. They correspond to the location $\sigma_k$ and the time $\tau_k$ at which vehicle $V_k$ ends the last delivery assigned to it up to time $t$.
The LKC, therefore, indicates when and where vehicle $V_k$ will become available again for another delivery.


The dispatching algorithm works in batches of time length $T_B$. It runs a new iteration at time instants $t\in\{hT_B,h\in\mathbb{Z}\}$. At the $h$-th iteration, it considers all the deliveries in the set
\begin{equation}
    \mathcal{D}_h = \{D_i\in\mathcal{D}: (h-1)T_B< t_o(i)\leq hT_B\},
\end{equation}
and assigns each of them to a vehicle while ensuring that the PUD of any delivery is lower than a predefined threshold value $\Delta$.

In order to do this, a weighted bipartite graph is constructed between the set $\mathcal{D}_h$ and the set $\mathcal{V}$ of all the vehicles. An edge between vehicle $V_j\in\mathcal{V}$ and delivery $D_i\in\mathcal{D}_h$ exists if the following condition is met
\begin{equation}
    \max(\tau_j,hT_B) + \Phi_t(\sigma_j,\ell_v(i)) \geq t_r(i) + \Delta.
    \label{condmatch}
\end{equation}
where $\max(\tau_j,hT_B) $ is the earliest time instant at which $V_j$ can start moving towards the pickup location, considering that it cannot do so before the current time instant $hT_B$.
The left-hand side of (\ref{condmatch}) represents the vehicle arrival instant at the pickup point, which cannot exceed the ready time $t_r(i)$ of delivery $D_i$ by more than $\Delta$, as prescribed by the delay constraint.
The same link is also assigned the weight $w_{i,j}$, defined as
\begin{equation}
    w_{i,j} = \Phi_d(\sigma_j,\ell_v(i)),
\end{equation}
corresponding to the distance that the vehicle must travel in order to reach the pickup location $\ell_v(i)$ of delivery $D_i$.

A minimum weight matching is then computed over the bipartite graph in polynomial time using the Hungarian algorithm. If the link between vehicle $V_j$ and delivery $D_i$ belongs to the matching, $D_i$ is assigned to $V_j$. Correspondingly, the LKC of $V_k$ are updated as
\begin{eqnarray}
 \tau_j & \leftarrow & \max\left[\max(\tau_j, hT_B) + \Phi_t(\sigma_j,\ell_v(i)), t_r(i)\right] + \Phi_t(\ell_v(i),\ell_c(i)); \label{uptime} \\
 \sigma_j & \leftarrow & \ell_c(i).
\end{eqnarray}
Expression (\ref{uptime}) can be explained as follows: vehicle $V_j$ starts moving towards the next pickup point as soon as possible, that is, immediately at $hT_B$ if it was idle, or at $\tau_j$ if it was serving another delivery.
It then takes $\Phi_t(\sigma_j,\ell_v(i))$ to get to the pickup point: however, if it arrives too early, it may need to wait until the time instant $t_r(i)$ before loading the new item, which explains the external $\max$ operation.
Once the item has been picked up, it then takes $\Phi_t(\ell_v(i), \ell_c(i))$ to deliver it at the customer location $\ell_c(i)$, which will become the next $\sigma_j$.
Upon assigning $D_i$ to $V_j$, the overall distance traveled to reach the pickup point and to deliver the item to the customer location is added to the total mileage traveled by vehicle $V_j$.

Since the cardinality $|\mathcal{V}|$ is in general different from $|\mathcal{D}_k|$, and since the bipartite graph is unlikely to be complete, it may happen that:
\begin{itemize}
    \item a vehicle $V_j$ is not assigned any trip: in this case, it simply keeps its LKC unaltered;
    \item a delivery $D_i$ is not assigned to any vehicle, mostly because there are no vehicles that can reach its pickup location in time. In this case, we \emph{generate} a new vehicle, adding it to the set $\mathcal{V}$, with LKC equal to $(\ell_v(i), t_r(i))$, and assign $D_i$ to it. 
    This is clearly an optimistic solution, which may lead to an underestimation of the overall mileage (the vehicle appears right at the desired pickup location $\ell_v(i)$).
    To address this issue, we associate a penalty term to the addition of a new vehicle: each new vehicle added to the system has a non zero starting mileage $M_s$, which is equal to the average mileage traveled by a vehicle between a delivery and the next pickup, as computed at the end of the simulation.
    This solution effectively mimics the fact that the new vehicle arrives at the pickup location $\ell_v(i)$ after having completed other services in the same urban area.
\end{itemize}
At the end of the last algorithm iteration, the overall traveled mileage is retrieved by summing the total distances traveled by each vehicle (including the initial distance $M_s$ for each of them).
The size of the required fleet is instead given by the cardinality of $\mathcal{V}$ at the end of the last batch.
The proposed algorithm ensures that all the items are delivered with a PUD lower than the threshold $\Delta$. The only scenario when this is not true is when the batch duration $T_B$ is higher than $T_p+\Delta$: in this case, the items ordered at the beginning of the batch cannot be picked up in time, since the algorithm iteration is performed when the delay constraint has already been violated. However, even in this case, the algorithm grants that the PUD is lower than $\Delta$ for all the remaining deliveries.

Extending the proposed algorithm to a scenario with bundling is straightforward. For each bundle, we first compute the time required to pick up and deliver all the bundled items (in a conveniently devised sequence), along with the total traveled mileage. Each bundle can then be represented as an equivalent single order: its pickup location corresponds to that of the first item, and its delivery location to that of the last item. Its ready time is defined as the ready time of the first item, while its effective maximum allowed pickup-to-delivery time (PUD) is computed to ensure that all items within the bundle are picked up within the original maximum allowed PUD $\Delta$.

\subsection*{Repositioning Strategy}
The dispatching algorithm outlined in the previous section can lead to fleet size overestimation if a proper repositioning strategy is not included. This occurs when a vehicle $V_j$ delivers an item to a location that is relatively far from all the vendor locations.
In this case, the time required to reach any other pickup point is higher than the allowed PUD threshold $\Delta$, making it impossible for $V_j$ to serve any other delivery.

To tackle this problem, we devised a repositioning strategy that associates each delivery $D_i$ with a \emph{Repositioning Return Location} (RRL) $\ell_r(i)$. This location is selected among the most popular vendor locations, and is the closest to $D_i$'s delivery location $\ell_c(i)$.
The repositioning strategy acts in two different ways, depending of the value of the travel time $\Phi_t(\ell_c(i), \ell_r(i))$:
\begin{itemize}
    \item if $\Phi_t(\ell_c(i), \ell_r(i)) > T_R$, where $T_R$ is a suitably chosen value, then the vehicle $V_j$ serving $D_i$ is sent back to the RRL $\ell_r(i)$ immediately after completing the delivery.
    As a matter of fact, in this case the delivery location $\ell_c(i)$ lies in an area far from the main customers, and it is unlikely that a pickup is required around it, so $V_j$ may remain stuck there for a long time;
    \item if $\Phi_t(\ell_c(i), \ell_r(i)) \leq T_R$, then the vehicle $V_j$ serving $D_i$ remains at the delivery location $\ell_c(i)$ waiting for new feasible deliveries.
    If, however, $V_j$ is not assigned any new delivery within a time interval equal to $T_W$, then it is sent to the RRL $\ell_r(i)$.
    In this case, the delivery location $\ell_c(i)$ is not too far from the most popular vendor locations: a new feasible delivery may enter the system, thus making an immediate repositioning unnecessary (and even detrimental from a mileage reduction perspective).
\end{itemize}
In any case, consecutive repositioning operations are not allowed: once a vehicle has been repositioned, it must wait a new delivery (potentially required at the same RRL that it reached with the repositioning).
The proposed strategy hence tries to find a balance between frequent repositioning operations (which lead to an increased overall mileage) and vehicles being stuck in faraway locations (which lead to an increased fleet size).

\subsection*{Theoretical Approximation}

Our theoretical approximation model formalizes how increased customer patience enables more efficient bundling of online grocery orders, ultimately reducing delivery mileage. We focus on the case in which at most two orders are bundled ($k=2$), since this dominates in practice in our datasets from Dubai.

We consider an order to be shareable when it can be bundled with at least one other order from the same vendor within the batching window. Under simplifying assumptions about vendor catchment areas and customer clustering (see Supplementary Material), we found that, for a vendor of popularity $\lambda$ (average order rate) and batch duration $\Delta$, the probability that an order is shareable can be expressed as
\begin{eqnarray}
P(\lambda, \Delta) &=& 1 - \frac{2}{\lambda \Delta}\left(e^{-\frac{\lambda\Delta}{2}} - e^{-\lambda\Delta}\right).
\end{eqnarray}

Since the batch duration is proportional to customer patience $\theta$ (with $\Delta = w\theta + z$), shareability rises monotonically with patience and approaches one as $\theta \to \infty$. Intuitively, popular vendors (large $\lambda$) achieve high shareability at much smaller patience levels.

At the city level, the relevant probability is obtained by averaging across the distribution of vendor popularities $f(\lambda)$. The posterior weighting gives
\begin{eqnarray}
P(\theta) &=& 1 - \frac{2}{\hat\lambda(w\theta+z)}
\int_0^\infty \left(e^{-\frac{\lambda(w\theta+z)}{2}} - e^{-\lambda(w\theta+z)}\right) f(\lambda) d\lambda
\end{eqnarray}
where $\hat\lambda$ is the mean vendor popularity. This expression links the aggregate probability of shareability directly to the distribution of vendor sizes in the market.

To evaluate the integral above, we approximate the vendor popularity distribution as bimodal, based on the observed pattern in the data: many small vendors following a power law, and a smaller set of large vendors following an exponential tail. The resulting cumulative distribution function is

\begin{eqnarray}
F(\lambda) =
\begin{cases}
1 - \dfrac{1}{(a\lambda+1)^b}, & 0 < \lambda \leq z_1 \\
1 - de^{-c\lambda}, & \lambda > z_2
\end{cases}
\end{eqnarray}
with continuity conditions linking $z_1$, $z_2$, and the parameters $(a,b,c,d)$. The posterior distribution for a random order is then
\begin{eqnarray}
\phi(\lambda) &=& \frac{1}{\mathbb{E}[\lambda]} \Bigg(\frac{ab\lambda}{(a\lambda+1)^{b+1}}\chi(0\leq \lambda \leq z_1)+cd\lambda e^{-c\lambda}\chi(\lambda > z_2)\Bigg),
\end{eqnarray}
where $\chi(\cdot)$ is an indicator function. Despite the presence of large vendors, the average popularity $\mathbb{E}[\lambda]$ remains low because of the dominance of small ones.

The fraction of bundled orders $F_B$ is necessarily smaller than $P(\theta)$, due to the limited maximum bundle size ($k=2$).
By approximating the impact of this parameter on the actual order pairing, we derive closed-form expressions for the probability $F_B(\theta)$ that two randomly drawn orders from the same vendor are bundleable.

Finally, the fraction of delivery mileage saved is expressed as
\begin{eqnarray}
F_{\text{dm}}(\theta) &=& F_B(\theta)\eta(\theta)
\end{eqnarray}
where $\eta$ is the expected proportion of redundant mileage eliminated when two trips are combined. Scaling this to the system-wide global mileage $F_{\text{gm}}(\theta)$ includes also the mileage saved in vehicle repositioning between subsequent deliveries, and accounts for the whole delivery activity across vendors.

In summary, the theoretical framework demonstrates that the expected savings from bundling are a sharply increasing function of customer patience, especially in markets with heterogeneous vendor sizes. Extending the shareability expression, we derive the fraction of total mileage savings $\Omega$ from bundling as a function of $\theta$ for the special case $k=2$. As illustrated, the theoretical curve well approximates the data across the entire considered span of the patience with significantly high $R^2$ values. For further details, please refer to Supplementary Note \ref{SI-note:TheoreticalModel}.

\subsection*{Life-Cycle Emission Estimation}

  We quantified life cycle emissions by evaluating both vehicle cycle and well-to-wheels (WTW) emissions for delivery fleets, including motorcycle fleet for Provider 1, and passenger vehicles and vans for Provider 2. We used the Greenhouse Gases, Regulated Emissions, and Energy Use in Transportation (GREET) model \cite{GREET} with Dubai-specific inputs for steel \cite{SteelEF} and gasoline for the evaluation \cite{ankathi2024well}. Life-cycle emissions were evaluated by Equation~\ref{Eq:emissions}.

  \begin{equation}
    E_{i,h,\ lifecycle} = \frac{1000 \times N_{i} \times E_{i, h, \ vehiclecycle}}{365 \times T_{i} } + E_{i,h, \ WTW} \times L_{i}
    \label{Eq:emissions}
  \end{equation}

\noindent
where $E_{i,h,lifecycle}$ is the life cycle emissions $h$ (\ce{CO2}/NOx/VOCs) of fleet $i$ (motorcycle/cars/vans), in unit g/day; $N_{i}$ is the number of vehicles in the fleet; $E_{i, h, \ vehiclecycle}$ is the vehicle cycle emissions $h$ of an individual vehicle, in unit kg; $T_{i}$ is the lifespan of an individual vehicle. We used 10 years for all fleets in this study \cite{zhao2024challenges}. $E_{i,h, \ WTW}$ is the WTW emissions $h$ in unit g/km; $L_{i}$ is the total vehicle miles traveled by the fleet, in unit km/day. Detailed emission parameters could be found in Supplementary Table \ref{SI-tab:EFregional}.

\section*{Acknowledgments}

The authors thank the Dubai Future Foundation and all the members of the MIT Senseable City Lab Consortium for supporting this research. The funders had no role in study design, data collection and analysis, decision to publish or preparation of the manuscript.

\section*{Author contributions}\label{sec6}

J.E., P.Z., F.L., G.R., P.S., M.M., A.K., S.F., and C.R. designed the research. J.E. and M.M. prepared the datasets. F.L., G.R., J.E., P.Z., A.K., S.F., and P.S. developed the methods. F.L. and G.R. performed the analysis. M.M. produced the figures. F.L., G.R., J.E., P.Z., S.F., M.M., and P.S. contributed to data interpretation. J.E., P.Z., F.L., G.R., P.S., A.K., and M.M. wrote the manuscript with the help of all the authors.


\section*{Competing interests}\label{sec7}
The authors declare no competing interests.

\section*{Inclusion and ethics}\label{sec8}

All authors have agreed to all manuscript contents, the author list and its order, and the author contribution statements. Any changes to the author list after submission will be subject to approval by all authors.

\bibliography{bundling}
\end{document}


\title{Supplementary Information for The Value of Patience in Online Grocery Shopping}

\maketitle

\section{Supplementary figures}

\begin{figure}[H]
    \centering
    \includegraphics[width=1.0\linewidth]{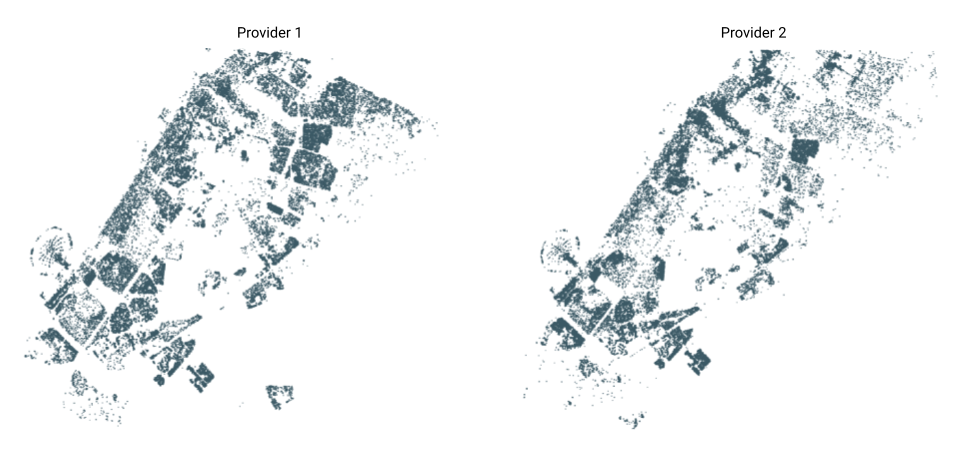}
    \caption{Spatial distribution of grocery orders for Data Providers 1 and 2. Each point represents the delivery location of an individual order, illustrating spatial variation in order density across the city. Differences between providers reflect distinct customer bases or service areas.}
    \label{fig:spatial_data_overview}
\end{figure}

\begin{figure}[H]
    \centering
    \includegraphics[width=1.0\linewidth]{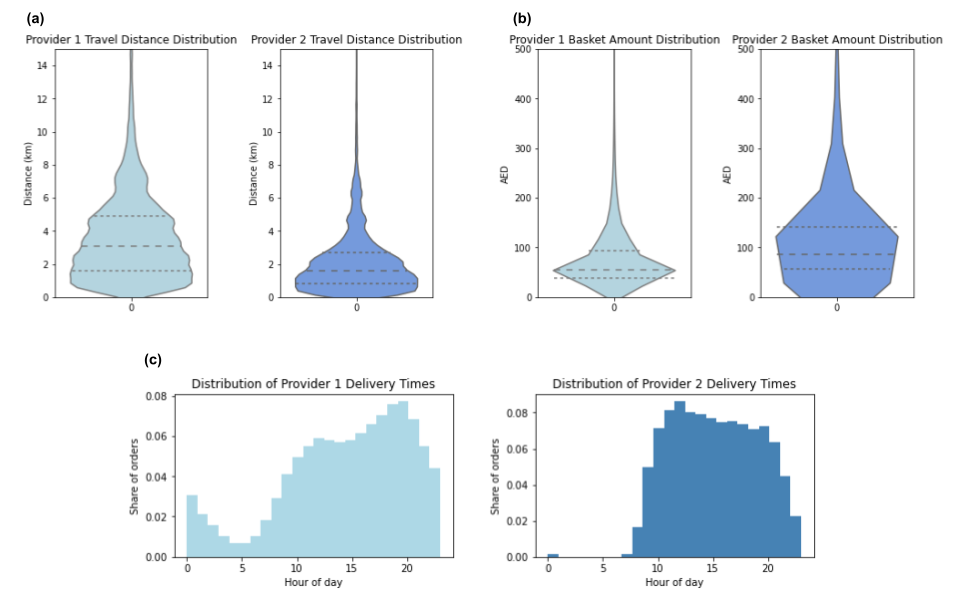}
    \caption{Comparison of the Provider 1 and Provider 2 datasets in terms of (a) travel distance distribution, (b) total value of goods in each basket (basket amount) distribution, and (c) delivery times distribution.}
    \label{fig:data_overview_graphs}
\end{figure}



\begin{figure}[H]
  \centering
  \begin{tikzpicture}[>=stealth]
  \def\base{-2.5};
  \def\lato{3.6};
  
  \begin{scope}
  
   \draw[black, thick, rounded corners] (-4,-3) rectangle +(8,6);
   
   \clip[rounded corners] (-4,-3) rectangle +(8,6);
   
   \node[] at (0,0) {\includegraphics[width=10cm]{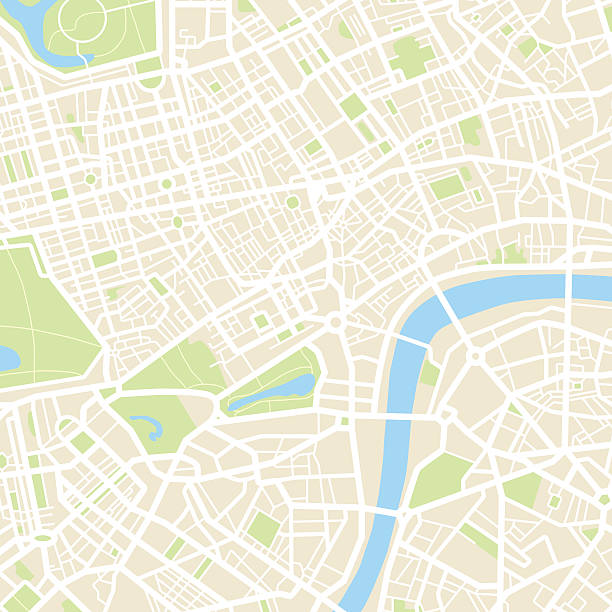}};
   
   \node[inner sep=0] (van1) at (-3.5, 2.4) {1\includegraphics[width=.5cm]{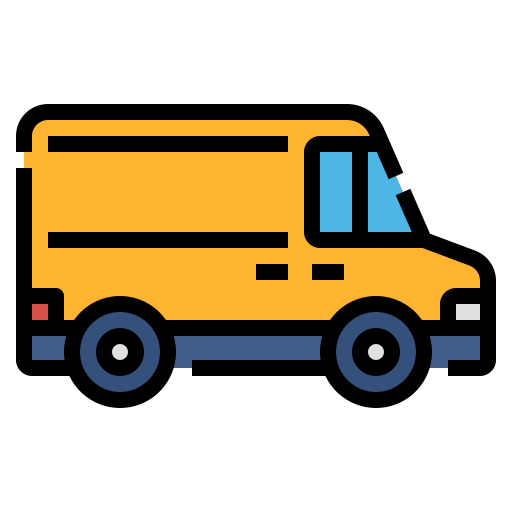}};
   
   \node[inner sep=0] (van2) at (-2.5, -0,2) {2\includegraphics[width=.5cm]{Figures/delivery.png}};
   
   \node[inner sep=0] (van3) at (2.5, 1.2) {3\includegraphics[width=.5cm]{Figures/delivery.png}};
   
   \node[inner sep=0] (van4) at (2.8, -2) {4\includegraphics[width=.5cm]{Figures/delivery.png}};
   
   \node[inner sep=0] (item1) at (-1, 2.5) {1\includegraphics[width=.5cm]{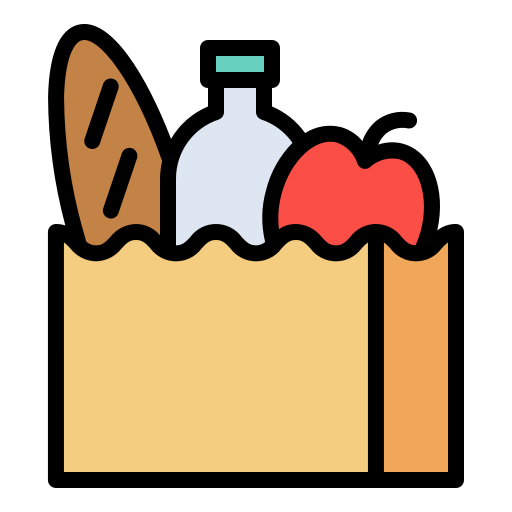}};
   
   \node[inner sep=0] (item2) at (1, -1.8) {2\includegraphics[width=.5cm]{Figures/food.png}};
   
   \node[inner sep=0] (item3) at (.2, .3) {3\includegraphics[width=.5cm]{Figures/food.png}};
   
   \node[inner sep=0] (item4) at (3.2, -.3) {4\includegraphics[width=.5cm]{Figures/food.png}};
   
   \node[inner sep=0] (item5) at (-2, -2.5) {5\includegraphics[width=.5cm]{Figures/food.png}};
   
   \draw[->,thick,green!80!black] (van1) to[bend left] (item1);
   
   \draw[->,thick,orange!80!black] (van1) to[bend left] (item3);
   
   \draw[->,thick,green!80!black] (van2) to[bend left] (item5);
   
   \draw[->,thick,orange!80!black] (van2) to[bend left] (item3);
   
   \draw[->,thick,green!80!black] (van3) to[bend right] (item3);
   
   \draw[->,thick,orange!80!black] (van3) to[bend right] (item1);
   
   \draw[->,thick,green!80!black] (van3) to[bend left] (item4);
   
   \draw[->,thick,green!80!black] (van4) to[bend right] (item2);
   
   \draw[->,thick,green!80!black] (van4) to[bend left] (item4);
   
  \end{scope}
  
  \filldraw[fill=blue!20, draw = black, thick, rounded corners] (\lato, \base) rectangle +(3,6);
  
  \draw[thick] (\lato+0.6, \base+1.2) -- (\lato+2.4, \base + 1);
  
  \draw[thick,red] (\lato+0.6, \base+1.2) -- (\lato+2.4, \base + 3);
  
  \draw[thick, red] (\lato+0.6, \base+2*1.2) -- (\lato+2.4, \base + 5);
  
  \draw[thick] (\lato+0.6, \base+2*1.2) -- (\lato+2.4, \base + 3);
  
  \draw[thick, red] (\lato+0.6, \base+3*1.2) -- (\lato+2.4, \base + 1);
  
  \draw[thick] (\lato+0.6, \base+3*1.2) -- (\lato+2.4, \base + 3);
   
  \draw[thick] (\lato+0.6, \base+3*1.2) -- (\lato+2.4, \base + 4);
  
  \draw[thick] (\lato+0.6, \base+4*1.2) -- (\lato+2.4, \base + 4);
  
  \draw[thick, red] (\lato+0.6, \base+4*1.2) -- (\lato+2.4, \base + 2);
  
  \foreach \y in {1,2,...,4}{
  \filldraw[fill=green!20, draw=black] (\lato+0.6, \base+1.2*\y) circle (.3cm);
  }
  
  \foreach \y in {1,2,...,5}{
  \filldraw[fill=red!20, draw=black] (\lato+2.4, \base+\y) circle (.3cm);
  }
  
  \node[inner sep=0] (vans) at (\lato+.6, \base +5.7) {Vehicle};
  
  \node[inner sep=0] (items) at (\lato+2.4, \base +5.7) {Item};
  
  \node[inner sep=0] (items) at (\lato+1.5, \base +.3) {\textbf{Matching}};
  
  \end{tikzpicture}
  \label{fig:matching}
  \caption{Snapshot of dispatching operations. At each time batch, based on the spatio-temporal coordinates of vehicles and ready items, a bipartite graph is constructed. An edge exists between a vehicle and an item if the former can reach the pickup location of the latter without violating the constraint on the maximum allowed pickup delay. A maximum (weighted) matching is computed to assign feasible deliveries to vehicles.}
  \end{figure}

\begin{figure}[H]
    \centering    \includegraphics[width=0.7\linewidth]{Figures/cluster radius k.jpg}
    \caption{The influence of cluster radius on relative mileage, for increasing values of $k$ based on Provider 2 orders.}
    \label{fig:cluster_radius}
\end{figure}

\begin{figure}[H]
    \centering
    \includegraphics[width=0.9\linewidth]{Figures/Fig_2-05.jpg}
    \caption{The influence of batch duration and PUD on mileage saving, fleet size, and total delivery time. Here, the maximum bundle size $k=4$.}
    \label{fig:results1}
\end{figure}

\begin{figure}[H]
    \centering
    \includegraphics[width=0.75\linewidth]{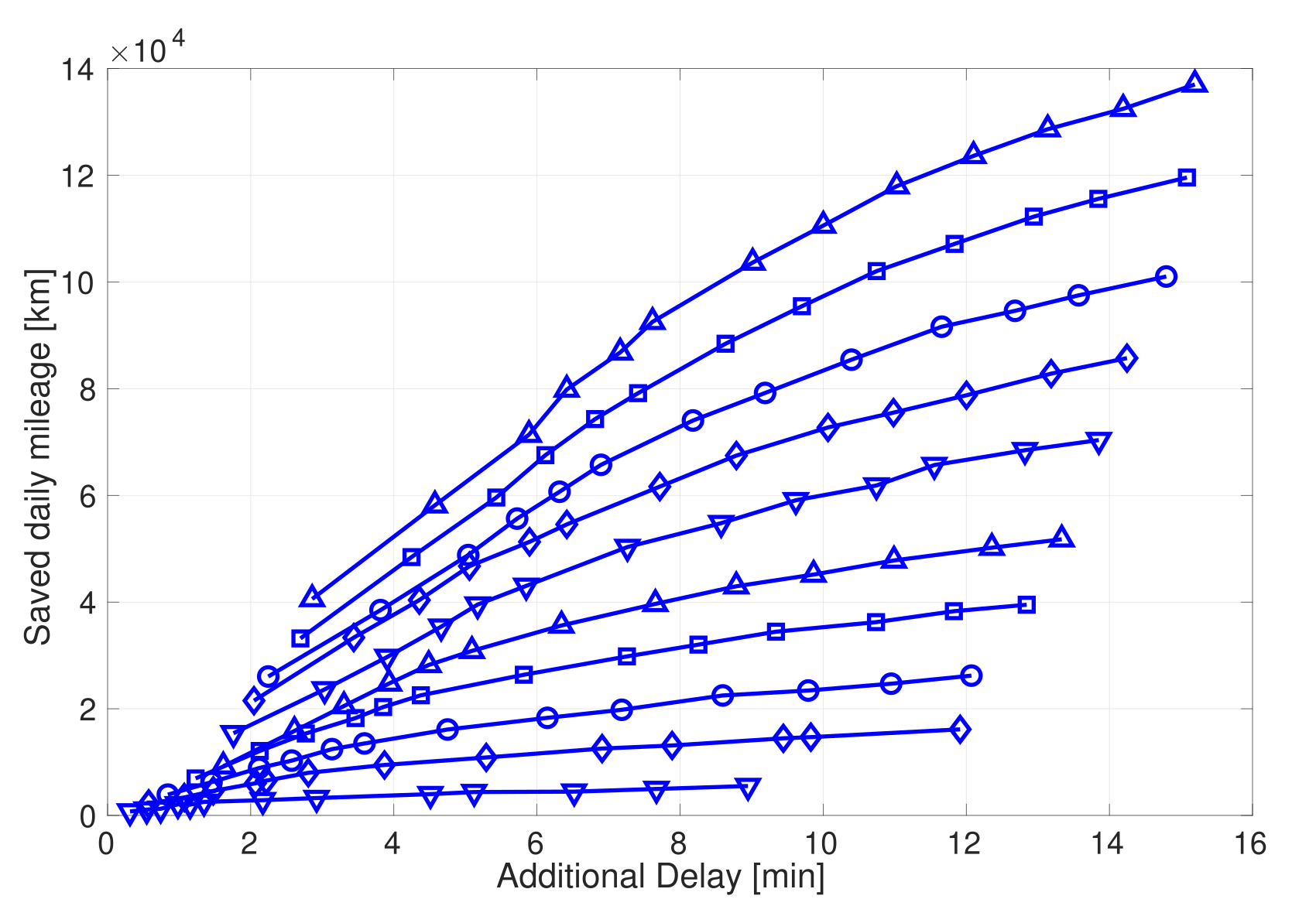}
    \caption{The relationship between additional delay and saving mileages with different trip densities. We simulated by sub-sampling through the orders in the day with the maximum number of orders. Lines from bottom to the top were simulated with random subsampled datasets of 10\%, 20\%, ..., 100\% trips from provider 1. }
    \label{fig:the_influence_of_trip_density}
\end{figure}

\begin{figure}[H]
    \centering
    \includegraphics[width=0.8\linewidth]{Figures/VARK_gain_mildel_PUDmil_2.png}
    \caption{The relationship between additional delay and saving mileages with different maximum bundling size $k$ for data provider 1. The scatter shows batch durations of 1 to 6 (by 1-minute intervals), and 8 to 20 minutes (by 2-minute intervals)}
    \label{fig:k_influence_delay_and_mileage}
\end{figure}

\begin{figure}[H]
    \centering
    \includegraphics[width=0.8\linewidth]{Figures/bundle size distribution contours.png}
    \caption{The contour plots illustrating the fraction of bundles with exactly two orders in them as a function of maximum bundle size and batch duration. As observed, for batch duration $<5$ minutes, the majority of bundles have size 2. Increasing both batch duration and and maximum bundle size increases the share of bundles with sizes greater than 2.}    \label{fig:bundle_size_contours}
\end{figure}

\begin{figure}[H]
    \centering
    \includegraphics[width=1.0\linewidth]{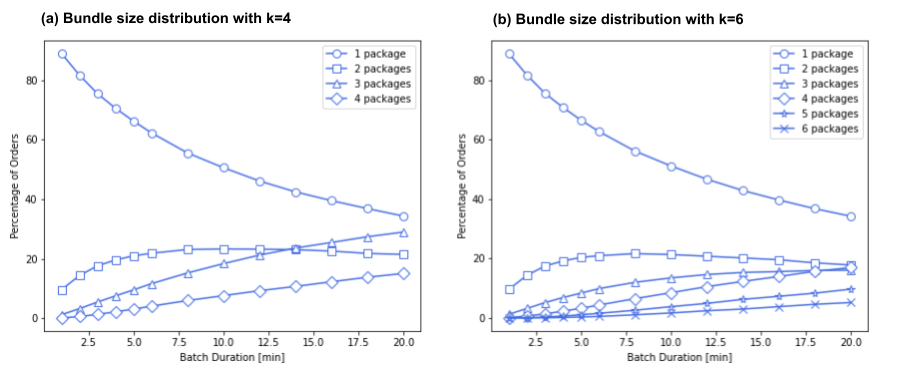}
    \caption{The percentage of bundles with $k\in\{1, 2, ..., K\}$ packages for $k=4$ and $k=6$. As observed, given small batch durations of less than 20 minutes, most of the bundles have a size of less than 4 packages. Hence, increasing $k$ beyond 4 without increasing batch duration does not result in significant changes.}
    \label{fig:bundle_size_distribution}
\end{figure}

\begin{figure}[H]
    \centering
    \includegraphics[width=0.8\linewidth]{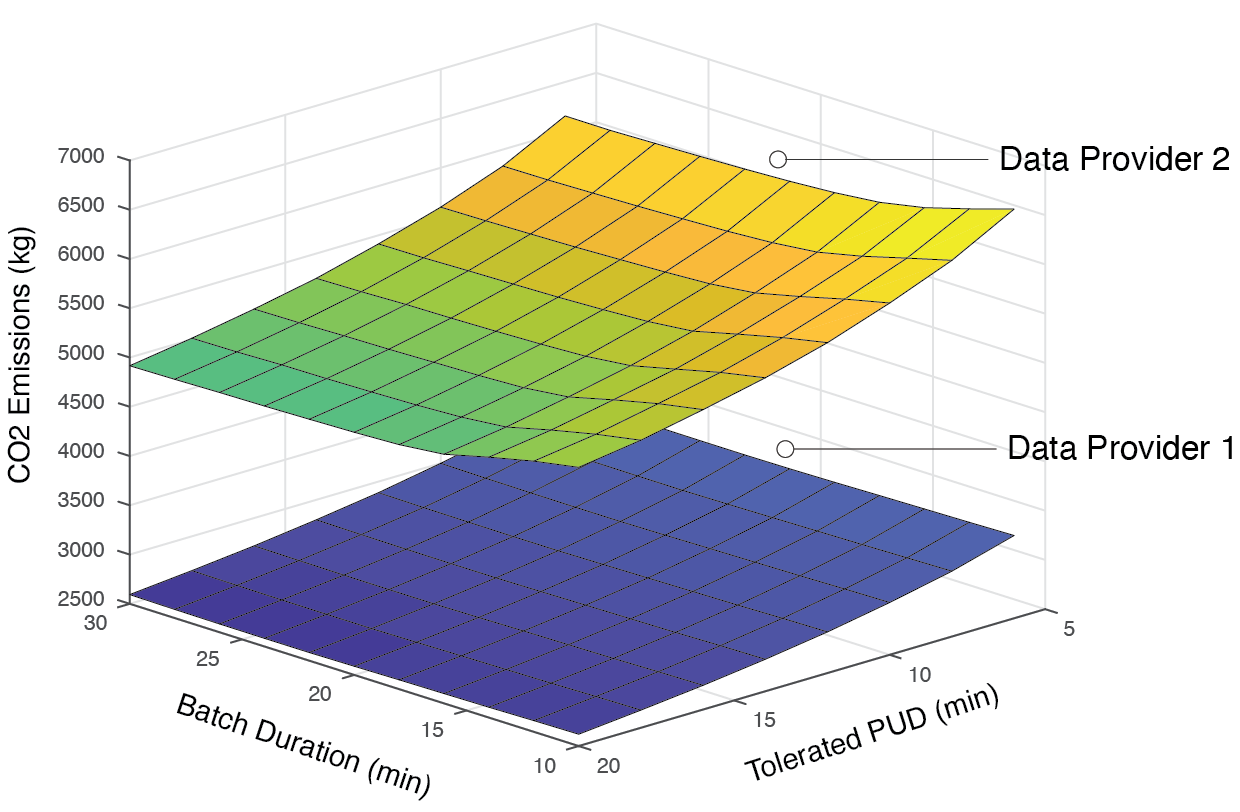}
    \caption{The influence of batch duration and maximum allowed pickup delay (PUD)  on \ce{CO2} emissions for both providers for $k=4$.}
    \label{fig:surface_emissions_fig}
\end{figure}

\begin{figure}[H]
    \centering
    \includegraphics[width=0.75\linewidth]{Figures/COMP_allCDFs.eps}
    \caption{Cumulative distribution function (CDF) of the computational time required for bundling and dispatching at the end of a batch.}
    \label{fig:CompTimeCDF}
\end{figure}

\begin{figure}[H]
    \centering
    \includegraphics[width=0.75\linewidth]{Figures/COMP_OrdTotTime2.eps}
    \caption{Observed relationship between the number of orders in a batch and the total computation time, including both bundling and dispatching, for the days with the maximum and the median number of orders.}
    \label{fig:Comp_OrdTotTime}
\end{figure}



\begin{figure}[H]
    \centering
    \includegraphics[width=0.75\linewidth]{Figures//FIN_NewEmi/FIN_gain_NOxdel_PUDemi_2.png}
    \caption{The average NOx emission savings with the additional delay. The PUD minimizing
emissions was used (22 min)}
    \label{fig:NOx_reduction}
\end{figure}

\begin{figure}[H]
    \centering
  \includegraphics[width=0.75\linewidth]{Figures//FIN_NewEmi/FIN_gain_VOCdel_PUDemi_2.png}
    \caption{The average VOCs emission savings with the additional delay. The PUD minimizing
emissions was used (22 min)}
    \label{fig:VOC_reduction}
\end{figure}

\section{Supplementary tables}

\begin{table}[H]
    \centering
    \caption{Overview of the two Providers datasets}
    \begin{tabular}{l|cc}
        \toprule
         & Provider 1 & Provider 2\\
        \midrule
        Number of Grocery Orders & 6,362,860 & 2,169,520 \\
        Number of Customers & 2,326,050 & 1,102,190 \\
        Number of Unique Vendors & 47 (1,455) & 41 \\
        Average Number of Orders per Customer & 2.7 & 1.97 \\
        Average Number of Orders per Vendor & 81,954 (4,373.1) & 52,915 \\
        \bottomrule
    \end{tabular}
    
    \label{tab:data_overview_table}
\end{table}

\begin{table}[H]
    \begin{center}
    \caption{Average emission factors for motorcycles, cars and vans \citep{SteelEF, GREET}}
    \begin{tabular}{ccccc}
    \toprule
    Stage & Emission & Motorcycle & Car & Van\\
    \midrule
    \multirow{3}{*}{WTW emission factors (g/km)}& \ce{CO2} & 58.2 & 129.3 & 323.3 \\
          &  NOx     & 0.0189 & 0.0420 & 0.105 \\
          &  VOCs    & 0.0216 & 0.0478 & 0.119 \\
    \midrule
    \multirow{3}{*}{Vehicle cycle emission factors (g/day)\footnotemark[1]}
          & \ce{CO2} & 159 & 907 & 1821 \\
          &  NOx     & 0.192 & 1.03 & 2.02 \\
          &  VOCs    & 0.200 & 0.772 & 1.44 \\
    \bottomrule
    \end{tabular}
    \label{tab:EFregional}
    \end{center}
    
\end{table}
\footnotetext[1]{Estimated by vehicle cycle emissions and their lifespan \citep{GREET}.}

\section{Supplementary notes}  

\subsection{Sensitivity analysis on the bundling and dispatching strategies}\label{note:SensitivityAnalysis}

Supplementary Figure \ref{fig:results1} illustrates the influence of batch duration and PUD on fleet size, relative mileage, and total delivery time when $k=4$. Extending batch duration provides more opportunities for bundling, effectively reducing fleet size and relative total mileage. For example, total mileage decreases by 8.8\% and 18.6\% when batch duration is extended to 10 and 18 minutes, respectively, compared to 6 minutes (PUD = 10 minutes). However, relative mileage is not highly sensitive to PUD. Across all batch duration values shown in Supplementary Figure \ref{fig:results1}(c), relative total mileage changes by no more than 0.5\% when PUD increases from 10 to 30 minutes. At the same time, increasing batch duration or PUD results in longer total delivery time. For instance, when PUD increases from 10 to 16 minutes (batch duration = 6 minutes), the total delivery time increases by 3.8 and 4.2 minutes for Providers 1 and 2, respectively.

Similar to mileage savings, life-cycle emission savings with bundling are more sensitive to batch duration than PUD, with both parameters positively influencing the emission reductions. As shown in Supplementary Figure \ref{fig:surface_emissions_fig}, increasing PUD from 10 to 30 minutes results in less than 1.2\% change in \ce{CO2} emission reductions across batch durations from 6 to 20 minutes. In contrast, increasing batch duration from 6 minutes to 10 and 18 minutes increases \ce{CO2} emission reductions by 9.3\% and 20.8\%, respectively. This is because vehicle travel related-WTW \ce{CO2} emissions, estimated from mileage, account for more than 80\% of life-cycle \ce{CO2} emissions without order bundling. Hence, life-cycle \ce{CO2} emission reductions are significantly influenced by mileage savings and inherit the parameter sensitivity.

As shown in Supplementary Figure \ref{fig:bundle_size_distribution}, relative mileage decreases with an increase in $r$. However, it eventually plateaus with further increases in the bundling radius. This is because $r$ not only determines the density of the order shareability network but also affects the efficiency of bundled deliveries by controlling the maximum distance between origins and destinations within a bundle. While increasing the bundling radius reduces mileage by consolidating delivery trips, it also increases intra-bundle travel mileage. Additionally, as previously mentioned, initial bundling cliques that satisfy the spatial-temporal proximity constraints are further partitioned into smaller cliques to comply with the maximum bundle size $k$. As a result, choosing an excessively large bundling radius may not significantly affect key outputs under the constraints of $k$ but would substantially increase computational costs due to the higher density of the order shareability network.

\subsection{Cost Estimation}\label{note:cost}
As discussed, by strategically bundling deliveries within the same geographic area, couriers can also minimize redundant travel, leading to lower operating and fixed costs. Here, we approximate the changes in the direct cost of delivery—excluding business operation costs—based on data from Provider 1 and publicly available datasets. This cost change is estimated as a function of fleet size ($N$) and total annual mileage ($M$), and incorporates parameters such as fleet purchase cost, labor cost, fuel cost, maintenance, and insurance, as modeled by equation \ref{eq: cost}:

\begin{equation}
\begin{split}
    Annual\_Cost &= N_{max} \times (P_{purchase} - P_{resale}) \div L  \\
        &+ N_{avg} \times c_w \\
        &+ f \times M \times c_f \\
        &+ N_{max} \times c_o
\end{split}
\label{eq: cost}
\end{equation}


\begin{table}[H]
    \centering
    \caption{Cost Estimation Parameters}
    \begin{tabular}{lc}
    \toprule
        \textbf{Variable} & \textbf{Selected Value} \\
        \midrule
        Car purchase price ($P_{purchase}$) & \$16400 \cite{NissanSunny2025}\\
        Motorcycle purchase price ($P_{purchase}$) & \$2200 \cite{HondaUnicorn2025}\\
        Car resale price ($P_{resale}$) & \$1800 \cite{Dubizzle2025}\\
        Motorcycle resale price ($P_{resale}$) & \$250 \cite{Dubizzle2025}\\
        Annual car maintenance, insurance, etc ($c_o$) &  \$370 \cite{CarCostCalculator2025} \\
        Annual motorcycle maintenance, insurance, etc ($c_o$) &  \$190 \cite{WaynesvilleCycle2025} \\
        Car fuel consumption ($f$) & 6.3 L/100km \cite{NissanSunny2025} \\
        Motorcycle fuel consumption($f$) & 2.5 L/100km \cite{HondaUnicorn2025}\\
        Annual labor wage and benefits $(c_w)$ & \$17,900 \cite{Dawson2025, KhaleejTimes2024}\footnotemark[1]\\
        Fuel price ($c_f$) & \$2.63 \cite{IPTEnergy2025} \\
        Vehicle Lifetime ($L$) & 10 years \\
     \bottomrule
    \end{tabular}
    
    \label{tab:costvars}
\end{table}

\footnotetext[1]{Estimations confirmed with our interviews.}

Following this formula and the proposed approach, we estimate that by introducing an additional 10-minute delay and bundling a maximum of four packages, the current direct annual cost of delivering grocery orders to courier companies would be reduced by 13\% if motorcycles are replaced by passenger cars due to higher volume of bundled orders, and 34\% if they continue delivering using motorcycles.

It is important to note that this cost function considers only the direct cost of delivery from the perspective of courier companies and does not account for broader societal impacts such as environmental costs, road usage, and traffic congestion, which are excluded from this analysis. Moreover, this approximation only considers the direct cost of delivery and does not include business operation costs such as software development, marketing, etc.

\subsection{Global and future projections of \ce{CO2} emission savings}\label{note:globalprojection}
We estimated the total \ce{CO2} emissions associated with grocery shopping by the share of grocery shopping in all trips. It is estimated that the \ce{CO2} emissions related to grocery shopping in the US is 17 million metric tons~\cite{USEPA2016}, equivalent to 5\% of the total emissions from passenger cars~\cite{USEPA2024}. From here, we estimated the annual \ce{CO2} savings from 5-minute delay from Equation \ref{eq:co2savings}:
\begin{equation}
    m_{\ce{CO2}} =  M_{\ce{CO2}} \times r_{gro} \times r_{online} \times s_{bundling}
\label{eq:co2savings}
\end{equation}
where $m_{\ce{CO2}}$ is the annual \ce{CO2} savings globally from 5-minute delay, in the unit of million tonnes (Mt); $M_{\ce{CO2}}$ is the global \ce{CO2} emissions of private cars and vans, estimated to be 3740 Mt ~\cite{IEA2023,IEA2023Cars}; $r_{gro}$ is the Share of grocery shopping in all trips, we take 5\% as before mentioned. $r_{online}$ is the share of global penetration rate of online grocery shopping; we take 18.1\% for 2023 and 26.3\% for 2028 \cite {Statistical2023onlineshare}; $s_{bundling}$ is the share of \ce{CO2} savings from 5-minute additional delay in Figure 4, 25.6\%.

Following the \ce{CO2} emissions savings, we estimated how many trees could absorb equivalent \ce{CO2} emissions and  social cost of carbon. \ce{CO2} emissions absorbed by one tree on average is 22 kg/year\cite{treeCO2}. Social cost of carbon were estimated based on the \ce{CO2} emissions savings and the mean unit estimate of social cost of carbon, \$185 per tonne of \ce{CO2} \cite{rennert2022comprehensive}. Global projection results in 2023 and 2028 could be found in Table \ref{tab:futureprojection}.

\begin{table}[H]
    \centering
    \caption{Global \ce{CO2} emission savings from grocery bundling annually}
    \begin{tabular}{lcc}
    \toprule
        \textbf{Variable} & \textbf{2023} & \textbf{2028}\\
        \midrule      
       
        \ce{CO2} savings from 5-minute delay (Mt) & 7.96 & 11.57\\
        Tree numbers equivalent\footnotemark[1] (million) & 366 & 531\\
        Social cost of carbon\footnotemark[2]  (billion USD) & 1.47 & 2.14\\
     \bottomrule
    \end{tabular}
    
    \label{tab:futureprojection}
\end{table}

\footnotetext[1]{Tree numbers equivalent to annual \ce{CO2} savings from 5-minute delay}
\footnotetext[2]{Social cost of carbon of annual \ce{CO2} savings from 5-minute delay}

\subsection{Computational Time}\label{note:ComputationTime}
A real-time implementation of the proposed framework inherently requires that bundling and dispatching operations, computed at the end of every batch, must be completed in a reasonably short time. To shed light on this aspect, we measured the overall time requested to perform both tasks for each batch, considering the day with the highest number of orders within the dataset. The algorithms were implemented in C++ and executed on a Linux workstation equipped with an Intel Core i7-3930K CPU and 16\,GB of RAM.
As shown in Supplementary Figure \ref{fig:CompTimeCDF}, the required time was always below 3\,s, with only minimal discrepancies between $k=4$ and $k=6$. 
In particular, as the maximum bundle size $k$ increases, more time is required for the bundling algorithm, since larger cliques are to be checked, and the optimal pickup and deliver order within each one must be determined.
Conversely, the dispatching time decreases, since orders are grouped into a smaller set of bundles, that can be consequently assigned faster.
Both algorithms are run at the end of each batch, upon collecting information about all the orders issued during the batch, as well as about the locations of the available vehicles.
The computation time for the bundling grows almost linearly with the number of orders in a batch duration, with a slope that depends on (and increases with) the maximum bundle size $k$.
The computation time for the dispatching instead depends on both the number of orders in the batch and the number of vehicles. Nevertheless, in a realistic deployment where the number of vehicles is either fixed or proportional to the expected amount of orders, the main factor that impacts the computation time is again the number of orders.
Overall, the total computation time (bundling plus dispatching) was observed to grow linearly with the number of orders in a batch (correlation coefficient higher than 0.99), with an increase of about 1\,s every 267 orders, as depicted in Supplementary Figure~\ref{fig:Comp_OrdTotTime}.
The largest fraction of this time is actually spent in determining trip durations and lengths. Considering that such operations can be considerably optimized over time by properly leveraging previous measurements (e.g., through machine learning approaches), the resulting computational time confirms the feasibility of a real-time implementation of the proposed framework.

\subsection{Theoretical Model}\label{note:TheoreticalModel}
\subsubsection{Order Shareability Probability vs Customer Patience}
 We consider an order \emph{shareable} when it can be potentially bundled with at least one other order from the same vendor.
 This probability can be expressed as
 \begin{equation}
  P(\lambda, \Delta) = 1 - \frac{2}{\lambda\Delta}e^{-\frac{\lambda\Delta}{2}}\left(1-e^{-\frac{\lambda\Delta}{2}}\right),
  \label{iniexp}
 \end{equation}
 where $\Delta$ is the batch duration and $\lambda$ is the popularity of the vendor (that is, the average number of orders departing from that vendor per second).
 Replacing the batch duration with the customer patience $\theta$, which is linearly related to $\Delta$ as
 \begin{equation}
  \Delta = w\theta + z,
 \end{equation}
 we find
 \begin{equation}
  P(\lambda, \theta) = 1 - \frac{2}{\lambda(w\theta+z)}\left(e^{-\frac{\lambda(w\theta+z)}{2}}-e^{-\lambda(w\theta+z)}\right).
 \end{equation}
 This expression holds for a vendor with a known popularity. The curve starts from 0, is monotonically increasing and approaches 1 as $\theta\rightarrow+\infty$.
 However, high values of the popularity $\lambda$ make the curve increase faster, since there are more orders departing from the same vendor in the same batch. 
 
 In order to find the average shareability probability in a given city, the curve must be averaged across the posterior vendor popularity distribution $\lambda f(\lambda)/\hat\lambda$, being $f(\lambda)$ the vendor popularity distribution of the considered city and $\hat\lambda$ the average vendor popularity.
 The general expression becomes
 \begin{eqnarray}
  P(\theta) & = & 1 - \frac{2}{\hat\lambda\left(w\theta+z\right)}\mathbb{E}_\lambda\left[e^{-\frac{\lambda(w\theta+z)}{2}}-e^{-\lambda(w\theta+z)}\right] \nonumber\\
  & = & 1 - \frac{2}{\hat\lambda\left(w\theta+z\right)}\int_0^\infty\left(e^{-\frac{\lambda(w\theta+z)}{2}}-e^{-\lambda(w\theta+z)}\right)f(\lambda)\de\lambda.
 \end{eqnarray}
 
 For the considered scenario in Dubai, the theoretical curve and the empirical curve are plotted in Figure~\ref{fig:shapat_math}.
 \begin{figure}
 \centering
 \includegraphics[width=\figw]{Figures/FigTheo/TEOOK_shapat_math.eps}
 \caption{Fraction of shareable orders as a function of the customer patience.}
 \label{fig:shapat_math}
\end{figure}

\subsubsection{Mathematical Derivation}
 The aim of this section is to derive a theoretical expression which expresses the fraction of mileage saving due to bundling as a function of the average customer patience. The expression is derived for the special case of $K=2$ (that is, at most two orders can be bundled together).

 The notation utilized throughout the paper is reported in Table \ref{tab:nota}, while the steps to derive the expression are:
 
 \begin{enumerate}
  \item Estimate Prior Popularity distribution $f(\lambda)$;
  \item Obtain Posterior Popularity distribution $\phi(\lambda)$;
  \item Compute fraction $F_S$ of shareable orders;
  \item Compute fraction $F_B$ of bundled orders;
  \item Derive fraction $F_\textrm{dm}$ of saved \emph{Delivery} Mileage;
  \item Derive fraction $F_\textrm{gm}$ of saved \emph{Global} Mileage;
  \item Rescale batch duration $\Delta$ to customer patience $\theta$.
 \end{enumerate}

 \begin{table}
  \centering
  \caption{Notation}\vspace{.2cm}
  \begin{tabular}{lc|lc}
   \hline\hline
   Quantity & Notation & Quantity & Notation\\
   \hline
   Order time of order $i$ & $t_o(i)$ & Vendor popularity & $\lambda$\\
   Vendor location of order $i$ & $\ell_v(i)$ & Fraction of shareable orders & $F_S$\\
   Customer location of order $i$ & $\ell_c(i)$ & Fraction of bundled orders & $F_B$\\
   Batch duration & $\Delta$ & Fraction of saved Delivery Mileage & $F_\textrm{dm}$ \\
   Max bundling distance & $\delta_c$ & Fraction of saved Global Mileage & $F_\textrm{gm}$ \\
   Radius of vendor served area & $U$ & Customer Patience & $\theta$\\
   \hline\hline
  \end{tabular}
  \label{tab:nota}
 \end{table}
 
 \vspace{3mm}
 \noindent\textbf{Prior Popularity Distribution}
 
 For a given vendor $S_i$, we define its \emph{popularity} $\lambda_i>0$ as the average number of orders from that vendor in a day. It can be equivalently seen as the \emph{intensity} of orders departing from $S_i$.
 The prior popularity distribution $f(\lambda)$ gives the probability that a randomly selected vendor has popularity $\lambda$.
 
 This distribution depends on the considered scenario, and can be retrieved from the available dataset.
 In our case, based on the orders placed across an entire week, we observed that the vendors can be grouped into two quite separate groups:
 \begin{itemize}
  \item a very large group with low to medium popularity, serving from less than 1 order/day to around 320 orders/day;
  \item a much smaller group of big vendors, serving more than 650 orders/day on average.
 \end{itemize}
 This lets us us define a bimodal popularity distribution $f(\lambda)$: the popularity of the large fraction of small vendors can be very well approximated by a power-law distribution; conversely, the popularity of the small group of big vendors seems to follow a more rapidly decreasing exponential distribution.
 This results in the following expression for the cumulative distribution function (CDF) $F(\lambda)$:
 \begin{equation}
  F(\lambda) = \left\{
               \begin{array}{ll}
                0 & \text{if } \lambda\leq0 \\
                \displaystyle 1- \frac{1}{(a\lambda+1)^b} & \text{if } 0<\lambda\leq z_1,\\
                \displaystyle 1- \frac{1}{(az_1+1)^b} & \text{if } z_1<\lambda\leq z_2,\\
                \displaystyle 1- de^{-c\lambda} & \text{if } \lambda>z_2,
               \end{array}
               \right.
 \label{popcdf}
 \end{equation}
 where the coefficients $a$, $b$, $c$ and $d$, as well as the cutoff value $z_1$, depend on the empirical distribution. Conversely, in order for $F(\lambda)$ to be continuous, it must hold
 \begin{equation}
  1- \frac{1}{(az_1+1)^b} = 1- de^{-cz_2},
 \end{equation}
 which leads to
 \begin{equation}
  z_2 = \frac{1}{c}\left(\ln(d) + b\ln(1+az_2)\right).
 \end{equation}
 Notice that $\fbig = 1/(az_1+1)^b$ can be seen as the fraction of \emph{big} vendors.
 
 We derived $a$ and $b$ by fitting the empirical complementary CDF (cCDF) in the interval $\lambda<0.22$ orders/day, while $c$ and $d$ were obtained by fitting the tail of the cCDF, that is, the interval $\lambda > 0.45$ orders/day. Their values are
 \begin{equation}
  a = 1002.039, \quad b = 0.925, \quad c = 4.283, \quad d = 0.061.
 \end{equation}
 Then, we set $z_1=0.1389$, corresponding to 200 orders/day, thus obtaining $z_2=0.4161$, corresponding to 600 orders/day. The empirical cCDF and the theoretic approximation are shown in Figure \ref{fig:popdist}.
 
 \begin{figure}
  \centering
  \includegraphics[width= \figw]{Figures/FigTheo/TEOOK_popdist.eps}
  \caption{Empirical popularity complementary CDF, obtained from a dataset over 7 days. The approximated theoretical cCDF is also plotted.}
  \label{fig:popdist}
 \end{figure}

 \vspace{3mm}
 \noindent\textbf{Posterior Popularity Distribution}
 
 The posterior popularity distribution $\phi(\lambda)$ is the probability that a randomly chosen order $t$ departs from a vendor with popularity $\lambda$.
 This can be computed using the prior popularity distribution as
 \begin{eqnarray}
  \phi(\lambda) & = & \sum_{i=1}^{N}\mathbb{P}\left[t\in \mathcal{S}_i, \lambda_i=\lambda\right] \nonumber \\
  & = & \sum_{i=1}^{N}\mathbb{P}\left[t\in \mathcal{S}_i| \lambda_i=\lambda\right]\mathbb{P}\left[ \lambda_i=\lambda\right] \nonumber \\
  & = & \sum_{i=1}^{N}\frac{\lambda}{\sum_{j=1}^N\lambda_j} f(\lambda),
  \label{compost}
 \end{eqnarray}
where $\mathcal{S}_i$ is the set of orders delivered from vendor $S_i$, whose popularity is $\lambda_i$, and $N$ is the total number of vendors. We also exploited the fact that the probability that an order is delivered from a vendor with a given popularity is given by the ratio between the number of orders delivered from that vendor and all the orders. This ratio is in turn equal to that between the popularity of the vendor and the sum of all the popularities.
The terms summed in (\ref{compost}) are all equal, since the popularity distribution is assumed to be the same for any vendor. Hence
\begin{eqnarray}
 \phi(\lambda) & = & N\frac{\lambda}{\sum_{j=1}^N\lambda_j} f(\lambda) \nonumber \\
 & = & \frac{\lambda f(\lambda)}{\mathbb{E}[\lambda]},
\end{eqnarray}
where $\mathbb{E}[\lambda] = \sum_{j=1}^N\lambda_j/N$ is the vendor average popularity. Notice that it can be also derived as $\int_0^\infty\lambda f(\lambda)\de\lambda$, which ensures that $\phi(\lambda)$ is correctly normalized.

In the considered scenario, taking the derivative of $F(\lambda)$ in (\ref{popcdf}) we get
\begin{equation}
 \phi(\lambda) = \frac{1}{\mathbb{E}[\lambda]}\left(\frac{ab\lambda}{(a\lambda+1)^{b+1}}\chi(0\leq\lambda\leq z_1) + cd\lambda e^{-c\lambda}\chi(\lambda>z_2)\right),
 \label{modphipdf}
\end{equation}
where $\chi(\cdot)$ is the indicator function, and
\begin{eqnarray}
 \mathbb{E}[\lambda] & = & \int_0^\infty\lambda f(\lambda)\de\lambda \nonumber \\
 & = & \frac{1}{a(b-1)} +\frac{\fbig}{a}\left(1 - \frac{b(az_1+1)}{b-1}\right)+ \frac{\fbig}{c}\left(1-\ln\left(\frac{\fbig}{d}\right)\right).
\end{eqnarray}
The resulting (still bimodal) pdf is depicted in Figure~\ref{fig:phipdf}. Despite the few big vendors collect thousands of orders, the mean is still low due to the large majority of the small vendors.

\begin{figure}
 \centering
 \begin{tikzpicture}
  \node at (0,0) {\includegraphics[width=\figw]{Figures/FigTheo/TEOOK_phipdf.eps}};
  
  \node[draw=red, fill=white] at (1,.5) {\includegraphics[width=7cm]{Figures/FigTheo/TEOOK_phipdf_zoom.eps}};
 \end{tikzpicture}
 \caption{The pdf $\phi(\lambda)$ of the popularity of the vendor of a randomly chosen order. A zoom on the low values is also highlighted.}
 \label{fig:phipdf}
\end{figure}

\subsubsection{Fraction of Shareable Orders}
\label{sec:frasha}
The four conditions that must be met for two orders $\alpha$ and $\beta$ to be potentially bundled are:
\begin{enumerate}
 \item the two orders are placed within the same batch duration $\Delta$, meaning that $t_o(\alpha)$ and $t_o(\beta)$ belong to the same batch;
 \item the locations of the two vendors, $\ell_v(\alpha)$ and $\ell_v(\beta)$, are closer than $\delta_v$;
 \item the locations of the two customers, $\ell_c(\alpha)$ and $\ell_c(\beta)$, are closer than $\delta_c$;
 \item the total length of the resulting combined trip is lower than the sum of the lengths of the two individual trips.
\end{enumerate}
However, simulations showed that, for $K=2$, the vast majority of the bundles (between 80\% and 90\%, depending on the batch duration $\Delta$) are formed by orders departing from the same vendor.
Therefore, we replace the second condition above with the condition $\ell_v(\alpha)=\ell_v(\beta)$.

In addition, we make the following assumptions about the spatio-temporal characteristics of the orders:
\begin{itemize}
 \item order times at vendor $S_i$ are generated according to a Poisson process of rate $\lambda_i$, being $\lambda_i$ the popularity of the vendor;
 \item the vendor and the customer locations, namely $\ell_v$ and $\ell_c$, of a generic order are modeled according to a bidimensional probability distribution $\rho_{v,c}(\ell_v,\ell_c)$ across the city area $\Omega$.
\end{itemize}

Call $P(\lambda,\Delta)$ the probability that a generic order $\alpha$, departing from vendor $S$ with popularity $\lambda$, located at $\mathbf{x}_{\alpha}$, and to be delivered at $\mathbf{y}_{\alpha}$, is potentially shareable with a random trip generated within the same batch of duration $\Delta$.
By assumption, this random trip must also depart from $S$.
The number of orders generated in a batch from $S$ is modeled as a Poisson random variable with parameter $\lambda\Delta$.
Therefore, we can write the probability $Q(\lambda,\Delta)=1-P(\lambda,\Delta)$ that $\alpha$ can \emph{not} be shared with any of these orders as
\begin{equation}
 Q(\lambda,\Delta) = \int_{\Omega}\int_{\Omega}\sum_{n=0}^\infty e^{-\lambda\Delta}\frac{(\lambda\Delta)^n}{n!}\left(1-p(\mathbf{x}_{\alpha},\mathbf{y}_{\alpha})\right)^n \rho_{x,y}(\mathbf{x}_{\alpha},\mathbf{y}_{\alpha}) \de\mathbf{y}_{\alpha} \de\mathbf{x}_{\alpha},
\end{equation}
where $p(\mathbf{x}_{\alpha},\mathbf{y}_{\alpha})$ is the probability that the order $\alpha$ can be potentially bundled with another order departing from the same vendor in the same batch.
The same equation can be rewritten as
\begin{equation}
 Q(\lambda,\Delta) = \int_{\Omega}\int_{\Omega}e^{-\lambda\Delta p(\mathbf{x}_{\alpha},\mathbf{y}_{\alpha})} \rho_{x,y}(\mathbf{x}_{\alpha},\mathbf{y}_{\alpha}) \de\mathbf{y}_{\alpha} \de\mathbf{x}_{\alpha}.
 \label{notshapro}
\end{equation}
In order to further develop (\ref{notshapro}), we need to make some assumptions on the spatial distribution $\rho_{x,y}(\mathbf{x}_{\alpha},\mathbf{y}_{\alpha})$. Vendors are likely to be spread all over the city area, in order to serve all the districts. Therefore, we consider a uniform distribution across the whole $\Omega$ for $\mathbf{x}_{\alpha}$.
Conversely, the customer location is strongly correlated with the vendor one. As a matter of fact, orders are delivered to the customer, in most cases, from the closest vendor location.
Hence, it is reasonable to assume that each vendor serves an area $\mathcal{U}(\mathbf{x}_{\alpha})$ around its location $\mathbf{x}_{\alpha}$.
Rewriting the joint distribution $\rho_{x,y}(\mathbf{x}_{\alpha},\mathbf{y}_{\alpha})$ as $\rho_{y|x}(\mathbf{y}_{\alpha}|\mathbf{x}_{\alpha})\rho_x(\mathbf{x}_{\alpha})$ yields
\begin{equation}
 Q(\lambda,\Delta) = \frac{1}{|\Omega|}\int_{\Omega}\int_{\mathcal{U}(\mathbf{x}_{\alpha})}e^{-\lambda\Delta p(\mathbf{x}_{\alpha},\mathbf{y}_{\alpha})} \rho_{y|x}(\mathbf{y}_{\alpha}|\mathbf{x}_{\alpha}) \de\mathbf{y}_{\alpha} \de\mathbf{x}_{\alpha},
 \label{notshapro2}
\end{equation}
where we leveraged the uniform distribution $\rho_x(\mathbf{x}_{\alpha}) = 1/|\Omega|$ of the vendors across the whole $\Omega$.

As to the conditional distribution $\rho_{y|x}(\mathbf{y}_{\alpha}|\mathbf{x}_{\alpha})$, in order to abstract from the peculiar topology of a given city, we assume that the customer location is uniformly deployed in the area $\mathcal{U}(\mathbf{x}_{\alpha})$, which is modeled as a circle of radius $U$ centered at $\mathbf{x}_{\alpha}$ (for any possible $\mathbf{x}_{\alpha}\in\Omega$).
Notice that $\rho_{y|x}(\mathbf{y}_{\alpha}|\mathbf{x}_{\alpha})$ hence depends only on the distance $|\mathbf{x}_{\alpha}-\mathbf{y}_{\alpha}|$ between the vendor and the customer location (that is, the length of the trip necessary to deliver the item).

The probability $p(\mathbf{x}_{\alpha},\mathbf{y}_{\alpha})$ that an order from $\mathbf{x}_{\alpha}$ to $\mathbf{y}_{\alpha}$ is potentially shareable with another order $\beta$ (within the same batch) from the same vendor location $\mathbf{x}_{\alpha}$ to a possibly different customer location $\mathbf{y}_{\beta}$ can be found as follows.

In principle, since order $\beta$ is from the same vendor, it should be $\mathbf{y}_{\beta}\in\mathcal{U}(\mathbf{x}_{\alpha})$.
However, an independent uniform distribution for $\mb{y}_\beta$ does not capture another important property of the customer locations, which are empirically found to exhibit a non negligible clusterization (see Section~\ref{sec:cluloc}).
In order to include this aspect while keeping a tractable approach, we assume that $\mb{y}_\beta$ is instead uniformly deployed within a circular region $\mathcal{U}_d(\mb{y}_\alpha)$ of radius $V$ centered at $\mb{y}_\alpha$.

The two remaining conditions for $\alpha$ and $\beta$ to be shareable are:
\begin{equation}
 |\mathbf{y}_\alpha - \mathbf{y}_\beta|\leq\delta_c,
 \label{cond1}
\end{equation}
\begin{equation}
 L_B < |\mathbf{x}_\alpha - \mathbf{y}_\alpha| + |\mathbf{x}_\alpha - \mathbf{y}_\beta|,
 \label{cond2}
\end{equation}
where $L_B$ is the total length of the route needed to deliver the two items when bundling is leveraged. The exact expression of $L_B$ depends on whether order $\alpha$ or order $\beta$ is delivered first. The optimal solution is to first deliver the closest one to the vendor location, and hence
\begin{equation}
 L_B = \min\left(|\textbf{x}_\alpha - \textbf{y}_\alpha|, |\textbf{x}_\alpha - \textbf{y}_\beta|\right) + |\textbf{y}_\alpha - \textbf{y}_\beta|.
 \label{defls}
\end{equation}
Figure~\ref{fig:routes} depicts a graphic sketch of the two possible routes for a bundle delivery.

\begin{figure}
 \centering
 \begin{tikzpicture}[>=stealth]
 \def\disto{6cm}
 \node[circle,minimum size =.3cm, fill=blue, inner sep = 0] (S) at (0,0) {};
 \node[above left = .01cm and .01cm of S] (labS.center) {$\mathbf{x}_\alpha$};
 
 \coordinate (desto) at ($(S) + (\disto,0)$);
 
 \fill[yellow] (desto) circle (2.5cm);
 
 \node[circle,minimum size =.3cm, fill=blue, inner sep = 0] (Da) at (desto) {};
 
 \node[above right = .01cm and .01cm of Da] (labDa.center) {$\mathbf{y}_\alpha$};
 
 \node[circle,minimum size =.3cm, fill=green, inner sep = 0, above left = 1cm and 1cm of desto] (Db1) {};
 \node[above left = .01cm and .01cm of Db1] (labDb1.center) {$\mathbf{y}_\beta$};
 
 \node[circle,minimum size =.3cm, fill=purple, inner sep = 0, below right = .8cm and 1.5cm of desto] (Db2) {};
 \node[above right = .01cm and .01cm of Db2] (labDb2.center) {$\mathbf{y}_\beta$};

 \draw[green!50!black,->,thick] (S) -- (Db1) -- (Da);
 \draw[red!50!black,->,thick] (S) -- (Da) -- (Db2);
 \draw (desto) -- node[midway,xshift=.2cm,yshift=-.2cm] {$\delta_c$} +(-130:2.5);
 
 \draw[dashed, thick] (-30:\disto) arc (-30:30:\disto); 
 
 
 \end{tikzpicture}
 \caption{Example of route for a bundle delivery. The vendor is the same for both items (located at $\mb{x}_\alpha$). The shaded area is the region $\mb{y}_\beta$ must belong to. However, if $\mb{y}_\beta$ is closer to $\mb{x}_\alpha$ than $\mb{y}_\alpha$, item $\beta$ is delivered first (green path), otherwise, item $\alpha$ is delivered first (red path). The dashed line represents the boundary: item $\beta$ is delivered first if and only if $\mb{y}_\beta$ lies to the left of it.}
 \label{fig:routes}
\end{figure}

Each condition can be mapped into a spatial region where $\mb{y}_\beta$ must lie in order for it to be met.
Condition (\ref{cond1}) simply identifies a circular area $\mathcal{C}$ centered on $\mb{y}_\alpha$, with radius $\delta_c$.
Condition (\ref{cond2}) can instead be reformulated based on the definition of $L_B$, and two cases can be distinguished:
\begin{itemize}
 \item if $|\mb{x}_\alpha - \mb{y}_\alpha| < |\mb{x}_\alpha - \mb{y}_\beta|$, meaning that item $\alpha$ is delivered first, we have $L_B =|\textbf{x}_\alpha - \textbf{y}_\alpha| + |\textbf{y}_\alpha - \textbf{y}_\beta|$, and condition (\ref{cond2}) becomes
 \begin{equation}
  |\mb{y}_\alpha - \mb{y}_\beta| < |\mb{x}_\alpha - \mb{y}_\beta|;
 \end{equation}
 \item if $|\mb{x}_\alpha - \mb{y}_\alpha| > |\mb{x}_\alpha - \mb{y}_\beta|$, meaning that item $\beta$ is delivered first, we have $L_B =|\textbf{x}_\alpha - \textbf{y}_\beta| + |\textbf{y}_\alpha - \textbf{y}_\beta|$, and condition (\ref{cond2}) becomes
 \begin{equation}
  |\mb{y}_\alpha - \mb{y}_\beta| < |\mb{x}_\alpha - \mb{y}_\alpha|.
 \end{equation}
\end{itemize}
Overall, the distance between the two customers must be lower than the length of the longest between the trips required to deliver the two orders. Calling $r$ the length $|\mb{x}_\alpha-\mb{y}_\alpha|$ of the trip to deliver order $\alpha$, the two cases identify two disjoint regions of the plane, namely
\begin{eqnarray}
 \mathcal{A} & = & \left\{\mb{y}: |\mb{y}-\mb{x}_\alpha| > r \wedge |\mb{y}-\mb{y}_\alpha| < |\mb{y}-\mb{x}_\alpha| \right\}, \\
 \mathcal{B} & = & \left\{\mb{y}: |\mb{y}-\mb{x}_\alpha| < r \wedge |\mb{y}-\mb{y}_\alpha| < r\right\},
\end{eqnarray}
such that condition (\ref{cond2}) is met if and only if $\mb{y}_b\in\mathcal{A}\cup\mathcal{B}$. The two regions are graphically shown in Figure~\ref{fig:regions}.
\begin{figure}
 \centering
 \begin{tikzpicture}
  \def\dist{2cm}
  
  \coordinate (A) at (0,0);
  \coordinate (B) at (\dist,0);
  \coordinate (H) at (.5*\dist,0);
  
  \fill[yellow] (B) circle (\dist);
  
  \fill[cyan] ($(H) + (0,-1.2*\dist)$) rectangle +(1.6*\dist,2.4*\dist);
  
  \draw[thick] ($(H) + (0,-1.2*\dist)$) to +(0,2.4*\dist);

  \begin{scope}
   \clip (B) circle (\dist);
   \fill[yellow] (A) circle (\dist);
  \end{scope}
  
  \draw ($(A)+(60:\dist)$) arc (120:240:\dist);
  \draw ($(A)+(60:\dist)$) arc (60:-60:\dist);

  \node[circle,minimum size =.2cm, fill=blue, inner sep = 0] (S) at (A) {};
  \node[above left = .01cm and .01cm of S] (labS.center) {$\mathbf{x}_\alpha$};
  
  \node[circle,minimum size =.2cm, fill=blue, inner sep = 0] (D) at (B) {};
  \node[above right = .01cm and .01cm of D] (labD.center) {$\mathbf{y}_\alpha$};
  
  \draw[thick,red] (A) to node[above,xshift=.2cm] {$r$} (B);
  
  \node[font = \Large, above = .6cm of H] {$\mathcal{B}$};
  \node[font = \Large, below right = .6cm and .7cm of B] {$\mathcal{A}$};
  
 \end{tikzpicture}
 \caption{Graphic representation of the two plane regions $\mathcal{A}$ and $\mathcal{B}$ where the customer of order $\beta$ must lie in order to meet condition (\ref{cond1}).}
 \label{fig:regions}
\end{figure}
The two orders $\alpha$ and $\beta$ are shareable only if both (\ref{cond1}) and (\ref{cond2}) are met, that is, if $\mb{y}_\beta\in\left(\mathcal{A}\cup\mathcal{B}\right)\cap\mathcal{C}$.

The exact expression of this resulting region depends on the value of $r$. We can distinguish two cases:
\begin{itemize}
 \item if $\delta_c<r$, it can be easily inferred from Figure~\ref{fig:regions} that $\mathcal{C}\subset\mathcal{A}\cup\mathcal{B}$. Therefore, condition (\ref{cond1}) is sufficient, and the two orders $\alpha$ and $\beta$ are shareable if $\mb{y}_\beta$ lies in a circle of radius $\delta_c$ centered at $\mb{y}_\alpha$;
 \item if $\delta_c>r$, the shape of the resulting region has a more complex shape, as shown in Figure~\ref{fig:apprarea} (shaded plus striped area). The closed form expression for the area of this region can be obtained, but is not suitable for subsequent mathematical derivations.
 Thus, we look for a lower bound of the shareable probability by replacing it with a simpler shape given by the union of two semicircles, centered at $\mb{y}_\alpha$, of radii $r$ and $\delta_c$. In Figure~\ref{fig:regions}, this corresponds to removing the two small striped regions.
 The resulting approximation error is quantified in subsection (\ref{sec:approerro}).
\end{itemize}

\begin{figure}
 \centering
 \begin{tikzpicture}
  \def\delc{3cm}
  \def\dist{2.5cm}
  
  \coordinate (A) at (0,0);
  \coordinate (B) at (\dist,0);
  \coordinate (H) at (.5*\dist,0);
  
  \begin{scope}
   \clip ($(H) + (0,-\delc)$) rectangle +(2*\delc,2*\delc);
   \fill[fill=cyan,postaction={pattern={Lines[angle=45,distance={6pt},line width=3pt]},pattern color=yellow}] (B) circle (\delc);
  \end{scope}
  
  \begin{scope}
   \clip ($(B) + (0,-\delc)$) rectangle +(\delc,2*\delc);
   \fill[fill=yellow] (B) circle (\delc);
  \end{scope}
  
  \fill[fill=yellow] (B) circle (\dist);
  
  \begin{scope}
   \clip ($(H) + (0,-1.1*\delc)$) rectangle +(2*\delc,2.2*\delc);
   \draw[thick] (B) circle (\delc);
  \end{scope}
  
  \begin{scope}
   \clip ($(B) + (0,-\delc)$) rectangle +(-\delc,2*\delc);
   \draw[thick] (B) circle (\dist);
  \end{scope}
  
  \draw[thick] (A) circle (\dist);
  
  \draw[dashed] ($(B) + (0,\delc)$) to +(0,-2*\delc);
  \draw[dashed] ($(H) + (0,\delc)$) to +(0,-2*\delc);
  
  \node[circle,minimum size =.2cm, fill=blue, inner sep = 0] (S) at (A) {};
  \node[above left = .01cm and .01cm of S] (labS.center) {$\mathbf{x}_\alpha$};
  
  \node[circle,minimum size =.2cm, fill=blue, inner sep = 0] (D) at (B) {};
  \node[above right = .01cm and .01cm of D] (labD.center) {$\mathbf{y}_\alpha$};
  
  \draw[thick,red] (A) to node[above,xshift=.2cm] {$r$} (B);
  
  \draw[thick,red] (B) to node[above, xshift=.1cm] {$\delta_c$} +(-60:\delc);
  
 \end{tikzpicture}
 \caption{Graphic representation of the region where $\mb{y}_\beta$ must lie in order for orders $\alpha$ and $\beta$ to be bundled (colored area), when $r<\delta_c$. The two small striped areas can be excluded for the sake of approximating the region with a more tractable shape.}
 \label{fig:apprarea}
\end{figure}

The probability $p(\mb{x}_\alpha,\mb{y}_\alpha)$ is then computed as the ratio between the area of $\left(\mathcal{A}\cup\mathcal{B}\right)\cap\mathcal{C}$ and the area $\pi V^2$ of $\mathcal{U}_d(\mb{y}_\alpha)$.

Puttig together the geometric considerations yields
\begin{equation}
 p(\mb{x}_\alpha,\mb{y}_\alpha) = p(r) = \left\{
 \begin{array}{ll}
  \displaystyle \frac{\delta_c^2}{V^2} & \text{if } r>\delta_c, \\
  \displaystyle \frac{r^2+\delta_c^2}{2V^2} & \text{if } r\leq\delta_c,
 \end{array}
 \right.
 \label{prosha}
\end{equation}
which depends only on the distance $r$ between $\mb{x}_\alpha$ and $\mb{y}_\alpha$.

Plugging this expression into (\ref{notshapro2}), and recalling that $\rho_{y|x}(\mb{y}_\alpha|\mb{x}_\alpha)$ is the distribution of $r$, namely
\begin{equation}
  f_r(x) = \frac{2x}{U^2} \quad\quad\text{ for } 0\leq x\leq U,
  \label{pdfr}
 \end{equation}
which is independent from $\mb{x}_\alpha$, we get
\begin{eqnarray}
 Q(\lambda,\Delta) & = & \int_0^Ue^{-\lambda\Delta p(r)} \frac{2r}{U^2} \de r, \\
 & = & \frac{2}{U^2}\int_0^{\delta_c}r\exp\left(-\frac{\lambda\Delta}{2V^2}(r^2+\delta_c^2)\right)\de r + \frac{2}{U^2}\int_{\delta_c}^Ur\exp\left(-\frac{\lambda\Delta\delta_c^2}{V^2}\right)\de r.
 \label{notshapro3}
\end{eqnarray}
By solving the integrals in (\ref{notshapro3}), and recalling that the probability that order $\alpha$ is shareable with at least another order is $P(\lambda,\Delta) = 1-Q(\lambda,\Delta)$, we can write $P(\lambda,\Delta)$ as
\begin{equation}
 P(\lambda,\Delta) = 1 - \left(1-\frac{\delta_c^2}{U^2}\right)e^{-\lambda\Delta\frac{\delta_c^2}{V^2}} - \frac{2}{\lambda\Delta}\left(\frac{V}{U}\right)^2
 e^{-\frac{\lambda\Delta}{2}\frac{\delta_c^2}{V^2}}\left(1-e^{-\frac{\lambda\Delta}{2}\frac{\delta_c^2}{V^2}}\right).
\end{equation}
Notice that if $U=\delta_c$, as was observed in the dataset, and $V=\delta_c$, which fits the data with high accuracy, the above equation greatly simplifies and yields (\ref{iniexp}).

The overall fraction $F_S$ of potentially shareable orders is hence obtained as
\begin{equation}
 F_S(\Delta) = \int_0^\infty P(\lambda,\Delta)\phi(\lambda)\de\lambda,
 \label{frasha}
\end{equation}
with $\phi(\lambda)$ expressed in (\ref{modphipdf}).

Figure~\ref{fig:sharbun} confirms the tightness of the proposed theoretical approximation. The dashed red curve has been obtained from the real dataset, computing the fraction of orders that can be bundled with at least one other order under the assumption that only orders from the same vendor can be bundled together.

\vspace{3mm}
\noindent\textbf{On the Clusterization of Customer Locations}
\label{sec:cluloc}
As per the rationale of the delivery system, we make the assumption that a vendor mostly serves customers located in its area.
Abstracting from the specific topology of the considered city, this justifies the choice, for the location of a generic customer served from a vendor located at $\mb{x}$, of a uniform distribution in a circle $\mathcal{U}(\mb{x})$ centered at $\mb{x}$, with radius $U$ that depends on the city area and on the vendor density.

However, it is also observed that customer locations tend to be clustered together: a vendor is likely to serve some specific areas (around its location) where customers are more concentrated.
We highlight this aspect in Figure~\ref{fig:clucoe}.
\begin{figure}
 \centering
 \includegraphics[width=\figw]{Figures/FigTheo/TEOOK_clucoe.eps}
 \caption{Clustering coefficient of the customers of the nine more popular vendors, compared with that of a uniform distribution.}
 \label{fig:clucoe}
\end{figure}
We identified the nine more popular vendors across a week in the dataset, and for each of them we first determined the locations of their served customers located within a given radius $U$. Then, these locations were utilized as vertices of a graph, and an edge was placed between each couple of vertices whose mutual distance is not greater than a Connection Radius $R$.
The clustering coefficient $\gamma$, defined as
\begin{equation}
 \gamma = \frac{2\sum_{i,j,k}\mb{A}_{ij}\mb{A}_{jk}\mb{A}_{ki}}{\sum_ik_i(k_i-1)},
\end{equation}
where $\mb{A}$ is the adjacency matrix of the graph and $k_i$ is the degree of the $i$--th node, was then computed with varying $R$ for each of the nine vendors.
For comparison, the same coefficient was calculated for a uniform distribution: in this case, the customer locations were randomly deployed in a circle of radius $U$ following a uniform distribution.

While Figure~\ref{fig:clucoe} indicates that, as expected, $\gamma$ grows with the Connection Radius $R$ for all the vendor locations, it also shows that the observed coefficient is always greater than that measured for customers deployed according to a uniform distribution.
This clearly confirms that customers tend to form clusters, a factor that increases the bundling probability but cannot be fully captured by assuming completely IID distributions around the vendor for the customers locations.

\vspace{3mm}
\noindent\textbf{Estimation of the Approximation Error}
\label{sec:approerro}
 The area of each of the striped areas in Figure~\ref{fig:regions} can be geometrically computed as
 \begin{equation}
  H = \frac{r}{4}\sqrt{\delta_c^2-\frac{r^2}{4}} + \frac{\delta_c^2}{2}\arctan\left(\frac{r/2}{\sqrt{\delta_c^2-r^2/4}}\right) - \frac{r^2}{2}\left(\frac{\sqrt{3}}{4} + \frac{\pi}{6}\right),
 \end{equation}
 and the resulting relative error hence is
 \begin{equation}
  \epsilon = \frac{2H}{\displaystyle 2H+\frac{\pi}{2}\left(r^2+\delta_c^2\right)},
 \end{equation}
 which can be rewritten as a function of the ratio $\eta=r/\delta_c$ as
 \begin{equation}
  \epsilon(\eta) = 1 - \frac{\pi(\eta^2+1)}{\eta\sqrt{1-\frac{\eta^2}{4}} + 2\arctan\left(\frac{\eta/2}{\sqrt{1-\eta^2/4}}\right) - 2\eta^2\left(\frac{\sqrt{3}}{4} + \frac{\pi}{6}\right)+\pi(\eta^2+1)},
 \end{equation}
 for $0\leq\eta\leq1$. This function is shown on the left panel of Figure~\ref{fig:anaerror}.
 
 \begin{figure}
  \centering
  \begin{tikzpicture}
   \node at (-3.7,0) {\includegraphics[width=7cm]{Figures/FigTheo/TEOOK_etaepsi.eps}};
   \node at (3.7,0) {\includegraphics[width=7cm]{Figures/FigTheo/TEOOK_avepsi.eps}};
  \end{tikzpicture}
  \caption{On the left, the theoretical approximation error as a function of the ratio $\eta$ between $r$ and $\delta_c$; on the right, the average approximation error $\bar\epsilon$ as a function of the ratio between the maximum delivery distance $U$ and $\delta_c$.}
  \label{fig:anaerror}
 \end{figure}

 According to the assumption that a customer is uniformly deployed within a circle of radius $U$ centered on the vendor location, it follows that the pdf of $r$ is given by (\ref{pdfr}), and the pdf of the ratio $\eta$ is therefore
 \begin{equation}
  f_\eta(x) = \frac{2\delta_c^2x}{U^2} \quad\quad\text{ for } 0\leq x\leq \frac{U}{\delta_c},
 \end{equation}
 and the average approximation error $\bar\epsilon = \mathbb{E[\epsilon(\eta)]}$ is given by
 \begin{equation}
  \bar\epsilon = 2\frac{\delta_c^2}{U^2}\int_0^{\min\left(\frac{U}{\delta_c},1\right)}x\epsilon(x)\de x,
 \end{equation}
 where we recall that $\epsilon(x)=0$ for $x>1$. Notice that the average error $\bar\epsilon$ depends on the ratio between $\delta_c$ and the maximum length of a delivery trip $U$.
 As long as $U>\delta_c$, the integral is constant, and $\bar\epsilon$ decreases with $U$; this is reasonable, since as $U$ becomes larger, it is less probable that $r<\delta_c$, and therefore that the approximation is needed at all.
 For $U<\delta_c$, as $U$ decreases the value of the integral decreases as well, but the constant factor $\delta_c/U$ increases.
 The value of $\bar\epsilon$ numerically computed as a function of $U/\delta_c$ is plotted in the right panel of Figure~\ref{fig:anaerror}. As can be observed, for $U\geq\delta_c$, which is confirmed by the data, the approximation error is not greater than 7\%. 

\subsubsection{Fraction of Bundled Orders}
Not all the orders that can be potentially bundled are actually bundled. From a theoretical viewpoint, finding the bundles is equivalent to find cliques in a graph $G=(\mathcal{V},\mathcal{E})$ whose vertices are orders and edges exist between couples of orders that can be bundled.

However, due to the maximum admissible bundle size $K$, cliques are to be first split into smaller subsets of cardinality at most $K$, and it may happen that a vertex belonging to a clique remains isolated, such that the corresponding order is therefore not bundled.

In a very general manner, we can obtain the fraction $F_B$ of bundled orders as
\begin{equation}
 F_B(\Delta) = (1-\psi_{n|b}(\Delta))F_S(\Delta),
 \label{bunfra}
\end{equation}
where $\psi_{n|b}(\Delta)$ is the probability that an order is not bundled given that it is potentially shareable.

We can give an expression of $\psi_{n|b}(\Delta)$ as follows. Consider a generic order $\alpha$ from a vendor with popularity $\lambda$, which is shareable with at least one other order from the same vendor. As per the assumptions above, the number of orders generated at the same vendor in a batch duration follows a Poisson distribution with parameter $\lambda\Delta$.
Of these orders, only a fraction is potentially shareable with $\alpha$, as expressed in (\ref{prosha}), depending on the distance $r$ between vendor and customer of the considered order.

By averaging over the distribution of $r$ in (\ref{pdfr}), we obtain that the equivalent intensity of potentially shareable orders with $\alpha$ is
\begin{eqnarray}
 \lambda^* & = & \lambda\left(\int_0^{\delta_c}\frac{r^2+\delta_c^2}{2V^2}\frac{2r}{V^2}\de r + \int_{\delta_c}^V\frac{\delta_c^2}{V^2}\frac{2r}{V^2}\de r\right) \nonumber \\
 & = & \lambda\left(\frac{\delta_c}{V}\right)^2\left[1-\frac{1}{4}\left(\frac{\delta_c}{V}\right)^2\right],
\end{eqnarray}
and hence, the pdf of the number $N$ of orders potentially shareable with $\alpha$, given that $\alpha$ can be shared with at least one other order, is
\begin{equation}
 f_N(n,\Delta) = \frac{e^{-\lambda^*\Delta}}{1-e^{-\lambda^*\Delta}}\frac{\left(\lambda^*\Delta\right)^n}{n!},
 \label{clisize}
\end{equation}
for $n\geq2$.

Under the (optimistic) assumption that all the orders potentially shareable with $\alpha$ are themselved mutually shareable with each other, (\ref{clisize}) gives also the distribution of the size of the clique of shareable orders $\alpha$ belongs to.
In the considered case $K=2$, only cliques with an even number of nodes can be split without leaving isolated nodes.
Conversely, the splitting of cliques with an odd number of nodes leaves one node isolated. Therefore, the probability that order $\alpha$ is not bundled, depending on the size of the clique it belongs to, can be modeled as
\begin{equation}
 \psi(n) = \left\{
 \begin{array}{ll}
  0 & \text{if $n$ is even} \\
  \frac{1}{n} & \text{if $n$ is odd}.
 \end{array}
 \right.
\end{equation}
By weigthing these value with the pdf in (\ref{clisize}) leads to the desired probability:
\begin{equation}
 \psi_{n|b}(\Delta) = C_b\sum_{n=2}^{\infty}\psi(n)f_N(n,\Delta).
\end{equation}
The term $C_b$ is a constant corrective term that must be introduced in order to take into account some aspects that are too involved to be explicitly modeled.
Firstly, not all the orders shareable with $\alpha$ are necessarily mutually shareable with each other. Secondly, one order may belong to more than one clique: in this case, the specific algorithm utilized determines the clique each node is assigned to, based on its target metric (minimize number of cliques, maximize their size...), and this may leave some nodes in cliques with even cardinality isolated as well.
Thirdly, the intensity of the shareable orders may vary if we allow orders from different vendors to be bundled toghether.
All these aspects influence the actual probability of bundling a giving order. However, the results show that all these effects can be effectively modeled by a constant term, which is retrieved from the data.

The fraction $F_B$ of bundled orders obtained from our theoretical approach is reported in Figure \ref{fig:sharbun}, together with the one measured from the data. The two curves appear to be very close across the entire span investigated of the batch duration $\Delta$.

\begin{figure}
 \centering
 \includegraphics[width=\figw]{Figures/FigTheo/TEOOK_sharbun.eps};
 \caption{Fraction $F_S$ of shareable orders (assuming bundling is possible only from same vendor) and fraction $F_B$ of observed bundled orders. Comparison between data and theoretical approach.}
 \label{fig:sharbun}
\end{figure}

\subsubsection{Fraction of Saved Delivery Mileage}
By construction, the bundling of two orders leads to a reduction of the overall mileage, as stated in the fourth condition in Section~\ref{sec:frasha}.
More precisely, the overall mileage $L_B$ from the pickup of the first order to the delivery of the last order must be lower than the sum of the lengths $L_i$, $i\in\{1,2,\ldots,K\}$ of the trips required to serve all the $K$ orders in the bundle separately.

This translates into a reduction of the overall delivery mileage, which can be computed as
\begin{equation}
 F_{\textrm{dm}}(\Delta) = \mu(\Delta) F_B(\Delta),
 \label{defdm}
\end{equation}
being $F_B$ the fraction of bundled orders introduced in (\ref{bunfra}), while $\mu(\Delta)$ is the average relative delivery mileage saving of a bundle of orders, that is,
\begin{equation}
 \mu(\Delta) = 1-\mathbb{E}\left[\frac{L_B}{\sum_{i=1}^KL_i}\right].
\end{equation}
The value of $\mu(\Delta)$ depends on the specific bundling algorithm, which may favor the minimization of the number of deliveries, the maximization of the mileage saving or other different metrics.
This value, retrieved from the available data, appears to be approximately linear with the batch duration, and can be expressed (see Figure~\ref{fig:linreg}) as
\begin{equation}
 \mu(\Delta) \simeq 0.0028\Delta + 0.2544.
\end{equation}

\subsubsection{Fraction of Saved Global Mileage}
The Global Mileage (GM) is given by the sum of the Service Mileage (SM) and the Delivery Mileage (DM).
The former includes all the miles traveled in \emph{service routes} (e.g., moving towards the pickup point of the next delivery, or repositioning the vehicle), while the latter corresponds to the mileage traveled while delivering items.
In the previous section, we computed the fraction $F_{\textrm{dm}}$ of saved DM as a function of $\Delta$.

Bundling orders together, however, has an impact also on the vehicle dispatching, and hence contributes to reduce the SM as well.
This reduction, and therefore the reduction of the GM, is much more involved to compute, since it depends on the employed dispatching algorithm.
However, the analysis of the routes completed by the serving vehicles reveals that the GM reduction is proportional to the one of the DM, that is
\begin{equation}
 F_{\textrm{gm}}(\Delta) = wF_{\textrm{dm}}(\Delta),
 \label{defgm}
\end{equation}
with $w = 1.325$. The fact that $w>1$ implies that bundling is able to reduce the SM even more than it does with the DM.
Bundling orders, indeed, removes up to $K-1$ service routes for each bundle. In the considered case $K=2$, one service route is avoided for each bundle, which helps explaining why the saved SM grows with the saved DM.

\begin{figure}
 \centering
 \includegraphics[width=0.8 \textwidth]{Figures/FigTheo/TEOOK_savmil.eps}
 \caption{Fraction of saved Delivery Mileage (DM) and Global Mileage (GM) as a function of the batch duration $\Delta$.}
 \label{fig:savmil}
\end{figure}

In Figure~\ref{fig:savmil}, the fraction $F_\textrm{dm}$ of saved DM and the fraction $F_\textrm{gm}$ of saved GM, computed as per (\ref{defdm}) and (\ref{defgm}), are shown and compared to those observed in the dataset.

\subsubsection{Saved Mileage as a function of Patience}
All the quantities derived in the previous sections are expressed in terms of the batch duration $\Delta$.
We now rescale them to find the relationship between the saved GM and the customer patience $\theta$.
It is straightforward to observe that $\theta$ must increase with $\Delta$, since longer batch durations necessarily imply, on average, longer waiting times for the customer.
From the simulation measurements, as shown in Figure~\ref{fig:linreg}, it is observed that a linear regression can effectively capture the relationship between these two quantities,
\begin{equation}
 \theta = 0.302\Delta + 0.677.
 \label{patbat}
\end{equation}

\begin{figure}
  \centering
  \begin{tikzpicture}
   \node at (-3.7,0) {\includegraphics[width=7.5cm]{Figures/FigTheo/TEOOK_reldelsav.eps}};
   \node at (3.7,0) {\includegraphics[width=7.5cm]{Figures/FigTheo/TEOOK_batch_pat.eps}};
  \end{tikzpicture}
  \caption{On the left, linear regression for the relationship between batch duration $\Delta$ and relative saved mileage $\mu$. On the right, linear regression for the relationship between Batch Duration $\Delta$ and customer patience $\theta$.}
  \label{fig:linreg}
 \end{figure}

By inverting the relationship (\ref{patbat}), we can obtain the batch duration as a function of the patience, and therefore write the fraction of saved Global Mileage as the function
\begin{equation}
 \Omega(\theta) = F_\textrm{gm}(3.311\theta-2.242).
\end{equation}

This function is compared with the real data in Figure~\ref{fig:savpat}. The theoretical curve well approximates the data across the entire considered span of the patience, between 1 and 7 minutes, with a calculated $R^2$ equal to 0.9915.

\begin{figure}
 \centering
 \includegraphics[width=.62\textwidth]{Figures/FigTheo/TEOOK_finfig.eps}
 \caption{Fraction of saved Global Mileage as a function of the customer patience.}
 \label{fig:savpat}
\end{figure}

\bibliography{bundling}